\documentclass[twocolumn]{aastex63}

\usepackage{amsmath}
\usepackage{booktabs}
\usepackage{threeparttable}
\usepackage{tabularx}

\submitjournal{ApJ}

\shorttitle{LIMFAST. II. High-z Galaxy Formation From Multi-Tracer LIM}
\shortauthors{Sun et al.}
\graphicspath{{./}{figures/}}
\begin{document}

\title{LIMFAST. II. Line Intensity Mapping as a Probe of High-Redshift Galaxy Formation}

\email{gsun@astro.caltech.edu}

\author{Guochao Sun}
\affiliation{California Institute of Technology, 1200 E. California Blvd., Pasadena, CA 91125, USA}

\author{Llu\'{i}s Mas-Ribas}
\affiliation{Jet Propulsion Laboratory, California Institute of Technology, 4800 Oak Grove Drive, Pasadena, CA 91109, USA}
\affiliation{California Institute of Technology, 1200 E. California Blvd., Pasadena, CA 91125, USA}

\author{Tzu-Ching Chang}
\affiliation{Jet Propulsion Laboratory, California Institute of Technology, 4800 Oak Grove Drive, Pasadena, CA 91109, USA}
\affiliation{California Institute of Technology, 1200 E. California Blvd., Pasadena, CA 91125, USA}

\author{Steven~R.~Furlanetto}
\affiliation{Department of Physics and Astronomy, University of California, Los Angeles, CA 90024, USA}

\author{Richard~H.~Mebane}
\affiliation{Department of Physics and Astronomy, University of California, Los Angeles, CA 90024, USA}
\affiliation{Department of Astronomy and Astrophysics, University of California, Santa Cruz, 1156 High Street, Santa Cruz, CA 95064, USA}

\author{Michael~O. Gonzalez}
\affiliation{California Institute of Technology, 1200 E. California Blvd., Pasadena, CA 91125, USA}

\author{Jasmine Parsons}
\affiliation{McGill University, Department of Physics \& McGill Space Institute, 3600 Rue University, Montr\'eal, QC, H3A 2T8}

\author{A.~C.~Trapp}
\affiliation{Department of Physics and Astronomy, University of California, Los Angeles, CA 90024, USA \\
\copyright 2022. All rights reserved.}

\begin{abstract}
The epoch of reionization (EoR) offers a unique window into the dawn of galaxy formation, through which high-redshift galaxies can be studied by observations of both themselves and their impact on the intergalactic medium. Line intensity mapping (LIM) promises to explore cosmic reionization and its driving sources by measuring intensity fluctuations of emission lines tracing the cosmic gas in varying phases. Using LIMFAST, a novel semi-numerical tool designed to self-consistently simulate LIM signals of multiple EoR probes, we investigate how building blocks of galaxy formation and evolution theory, such as feedback-regulated star formation and chemical enrichment, might be studied with multi-tracer LIM during the EoR. On galaxy scales, we show that the star formation law and the feedback associated with star formation can be indicated by both the shape and redshift evolution of LIM power spectra. For a baseline model of metal production that traces star formation, we find that lines highly sensitive to metallicity are generally better probes of galaxy formation models. On larger scales, we demonstrate that inferring ionized bubble sizes from cross-correlations between tracers of ionized and neutral gas requires a detailed understanding of the astrophysics that shape the line luminosity--halo mass relation. Despite various modeling and observational challenges, wide-area, multi-tracer LIM surveys will provide important high-redshift tests for the fundamentals of galaxy formation theory, especially the interplay between star formation and feedback by accessing statistically the entire low-mass population of galaxies as ideal laboratories, complementary to upcoming surveys of individual sources by new-generation telescopes. \\
\end{abstract}

% ========== Section 1: Introduction ========== %

\section{Introduction} \label{sec:intro}

The enormous amount of observational and modeling efforts over the past two decades have revealed an increasingly detailed and precise picture of the epoch of reionization (EoR). Following the onset of the first galaxy formation at $z>10$ \citep{Oesch_2016, Naidu_2020, Bouwens_2021, Harikane_2022} and being completed by $z\approx5$--6 \citep{McGreer_2015, Becker_2021, Cain_2021}, the neutral intergalactic medium (IGM) after cosmic recombination was ionized again by an accumulating background of energetic UV photons emerged from the evolving populations of early star-forming galaxies (\citealt{Fan2006,Stark2016ARA&A,DF2018PhR,Robertson2021arXiv}; but see also e.g., \citealt{Qin_2017} for alternative sources like quasars). 

An emerging technique in observational cosmology, line intensity mapping (LIM) has been widely recognized as a powerful method to study the co-evolution of galaxies and the IGM during the EoR \citep{Kovetz_2017, Chang_2019}. Particularly, the prospects of synergies among LIM surveys targeting at different (and usually complementary) tracers have attracted considerable attention in recent years, as more and more target lines being identified and observed at wavelengths across the electro-magnetic spectrum. Substantial theoretical effort has been made in recent years to investigate the scientific potentials of multi-tracer LIM. One important objective is to employ the large-scale complementarity between tracers of ionized and neutral regions in the IGM to tomographically measure the reionization history \cite[e.g.,][]{Lidz_2011, Gong_2012, Feng_2017, Heneka_2017, Dumitru_2019, Kannan_2022_LIM}. Such joint analyses can trace the growth of ionized regions and alleviate observational challenges like systematics and foreground contamination. Another major objective is to infer physical properties of the source population through simultaneous diagnosis of multiple spectral lines emitted from the multi-phase interstellar medium (ISM) and/or IGM \cite[e.g.,][]{Heneka_2017, Sun_2019, Yang_2021, SW_2021, Bethermin_2022arXiv}. Even though the coarse-grain averaged nature of these statistical measurements makes the interpretation challenging in many circumstances, these efforts have showcased the richness of astrophysical information about the EoR to be gleaned from multi-tracer LIM datasets. 

Nevertheless, the majority of modeling efforts in the LIM context can be broadly considered as ``semi-empirical'', which leverage a small number of simple, observationally-motivated trends to describe the source population and create mock LIM signals. Although these models provide quantitative expectations of LIM signals anchored to observations, clear physical connections among properties of the source population and different observables are often missing \cite[but see][who employ fully-detailed, radiation-magneto-hydrodynamic simulations of reionization with photoionization and radiative transfer modeling to study multi-tracer LIM]{Kannan_2022_THESAN, Kannan_2022_LIM}. Relatively little effort has been devoted so far to the development of physical yet efficient modeling frameworks that capture the essential astrophysical information, while being flexible enough to allow model testing and inference from multi-tracer LIM datasets. For these reasons, we have developed LIMFAST, a semi-numerical toolkit for self-consistently simulating a multitude of LIM signals during the EoR, as introduced in detail in Mas-Ribas et al. (2022, henceforth Paper~I). LIMFAST extends the 21cmFAST code, and implements significantly improved models of galaxy formation and line emission in the high-$z$ universe. 

In this work, we present a generalization and applications of the basic framework of LIMFAST introduced in Paper~I, by considering physically-motivated variations of stellar feedback and star formation law prescriptions. Given the important consequences these variations have on galaxy and IGM evolution during the EoR, we investigate their implications for a number of promising LIM targets for probing the EoR, including the 21 cm line of \ion{H}{1} and nebular lines at optical/UV (e.g., H$\alpha$, Ly$\alpha$) and far-infrared (e.g., [\ion{C}{2}], [\ion{O}{3}], CO) wavelengths. Such a generalization allows us to relate specific LIM observables to a fundamental picture of high-$z$ galaxy formation described by a balance maintained by star formation from the ISM and stellar feedback typically from supernovae \citep{Furlanetto_2017, Furlanetto_2021}. By characterizing how these astrophysical processes impact the summary statistics of different LIM signals, especially the auto- and cross-power spectra, we investigate how the underlying astrophysics of feedback-regulated star formation can be better understood from future LIM observations combining different line tracers. 

The remainder of this paper is structured as follows. In Section~\ref{sec:limfast}, we briefly review LIMFAST, including its basic code structure and functionalities. In Section~\ref{sec:models}, we specify key features of the galaxy formation model and its variations considered in this work, namely prescriptions for stellar feedback and the star formation law. We also introduce a supplement to the baseline nebula model discussed in Paper~I, which allows us to model lines emitted from the photodissociation regions (PDRs) and molecular gas irradiated by the interstellar radiation field sourced by star formation. In Section~\ref{sec:results}, we present the main quantitative results of this work about how variations of the galaxy model affect the reionization history, followed by how distinct forms of feedback and the star formation law may be revealed by multi-tracer LIM observations. We compare our results with previous work, discuss some limitations and caveats of the analyses presented, and outline several possible extensions of the current framework in Section~\ref{sec:discuss}, before summarizing our main conclusions in Section~\ref{sec:conclusions}. Throughout the paper, we assume cosmological parameters consistent with recent measurements by Planck Collaboration XIII \citep{Planck_XIII}. 

\section{The LIMFAST Code} \label{sec:limfast}
In Paper~I, we describe in detail the general features and applications of the LIMFAST code. Here, we only briefly review the key features of LIMFAST and refer interested readers to the paper for further details. 

Built on top of the 21cmFAST code \cite[][]{MFC_2011, Park_2019}, LIMFAST shares with it the efficient, semi-numerical configuration adopted to approximate the formation of the large-scale structure and the partitioning of mass into dark matter halos. Specifically, the evolution of density and velocity fields is calculated with the Lagrangian perturbation theory, whereas the hierarchical formation of structures and the growth of ionized regions are described by the excursion set formalism \cite[][]{LC_1993, MF_2007}, without resolving individual halos. Using the overdensity field derived, LIMFAST replaces the simplistic galaxy formation model used in 21cmFAST by a quasi-equilibrium model of high-$z$, star-forming galaxies introduced by \citet{Furlanetto_2017} and \citet{Furlanetto_2021}, which predicts a range of physical properties important for LIM studies, including the gas mass, stellar mass, star formation rate (SFR), metallicity, and so on. The line intensity fields of interest are then computed by integrating emissivities predicted by the photoionization and radiative transfer simulation code, \textsc{cloudy} \cite[][]{Ferland_2017}, over the halo mass function conditional on the local overdensity. Following \citet{MFC_2011}, we normalize integrals of the subgrid conditional halo mass function in the Press-Schechter formalism \citep{Bond_1991} to match the mean values from the Sheth-Tormen formalism.

LIMFAST coherently simulates a variety of LIM signals that trace the reionization and the underlying galaxy formation histories. In Paper~I, the simulated cosmic star formation rate density and the IGM neutral fraction evolution are verified by comparing against latest observations of high-$z$ galaxies/quasars and the cosmic microwave background (CMB), whereas LIM statistics of a suite of (mainly optical/UV) nebular lines typically from \ion{H}{2} regions (e.g., Ly$\alpha$, H$\alpha$, [\ion{O}{2}], and [\ion{O}{3}]) are compared with other high-$z$ model predictions in the literature. Extensions and variations of the baseline model presented in Paper~I, including an extended model of emission lines that predominantly originate from the neutral ISM (e.g., [\ion{C}{2}] and CO), are introduced in this work to facilitate the analysis.

\section{Models} \label{sec:models}

To understand the connection between astrophysics of galaxy formation and LIM signals originating from different environments, especially the multi-phase ISM, and demonstrate the astrophysical applications of multi-tracer LIM studies, we need to consider some plausible variations of the galaxy formation model, and preferably a large set of line signals sensitive to the variations and different gas phases. In what follows, we will describe how the baseline LIMFAST simulation presented in Paper~I is extended for such purposes.  

\subsection{Models of Galaxy Formation and Evolution} \label{sec:model-galaxy}

Following the galaxy formation model described in Paper~I based on \citet{Furlanetto_2017} and \citet{Furlanetto_2021}, the star formation and chemical evolution of individual halos can be described by a set of simple, coupled ordinary differential equations. 

Expressing any given mass in terms of the dimensionless quantity $\tilde{M} \equiv M/M_0$, where $M_0$ denotes the halo mass at some initial redshift $z_0$, and taking derivatives with respective to redshift for the time evolution (i.e., $\tilde{M}^\prime = d \tilde{M} / d z$), we can express the evolution of halo mass, gas mass, stellar mass, and metal mass as
\begin{equation}
\frac{\tilde{M}^\prime}{\tilde{M}} = -|\tilde{M}_0^\prime|~,
\label{eq:mh_prime}
\end{equation}

\begin{equation}
\frac{\tilde{M}_g^\prime}{\tilde{M}_g} = -|\tilde{M}_0^\prime|  \left[ X_g^{-1} - \frac{\epsilon(\mathcal{R}+\eta)}{|\tilde{M}_0^\prime| C_\mathrm{orb}} \left(\frac{1+z_0}{1+z}\right) \right]~,
\end{equation}

\begin{equation}
\tilde{M}_*^\prime = -\mathcal{R} \tilde{M}_g \left[ \frac{\epsilon}{C_\mathrm{orb}} \left(\frac{1+z_0}{1+z}\right) \right]~,
\label{eq:ms_prime}
\end{equation}

\begin{equation}
\frac{\tilde{M}_Z^\prime}{\tilde{M}_Z} = \mathcal{R} \left(-1 - \eta + y_Z Z^{-1} \right) \left[ \frac{\epsilon}{C_\mathrm{orb}} \left(\frac{1+z_0}{1+z}\right) \right]~,
\label{eq:ms_prime}
\end{equation}
where $C_{\rm orb}=(1+z_0)A_{\rm dyn} \Delta^{-1/2}_{\rm vir}$ characterizes the parameter dependence of the orbital timescale, which is set by some normalization factor $A_{\rm dyn}$ and the virial overdensity of a collapsed halo $\Delta_{\rm vir} = 18\pi^2$. A constant return fraction $\mathcal{R}=0.25$ is taken to describe the fraction of stellar mass recycled back to the star-forming gas due to stellar evolution \citep{Benson_2010PhR, Tacchella_2018}. A metal yield $y_Z=0.03$ is adopted, which is appropriate for metal-poor, Population II (Pop~II) stars with a typical initial mass function \citep{Benson_2010PhR}. 

\begin{figure}[!ht]
 \centering
 \includegraphics[width=0.47\textwidth]{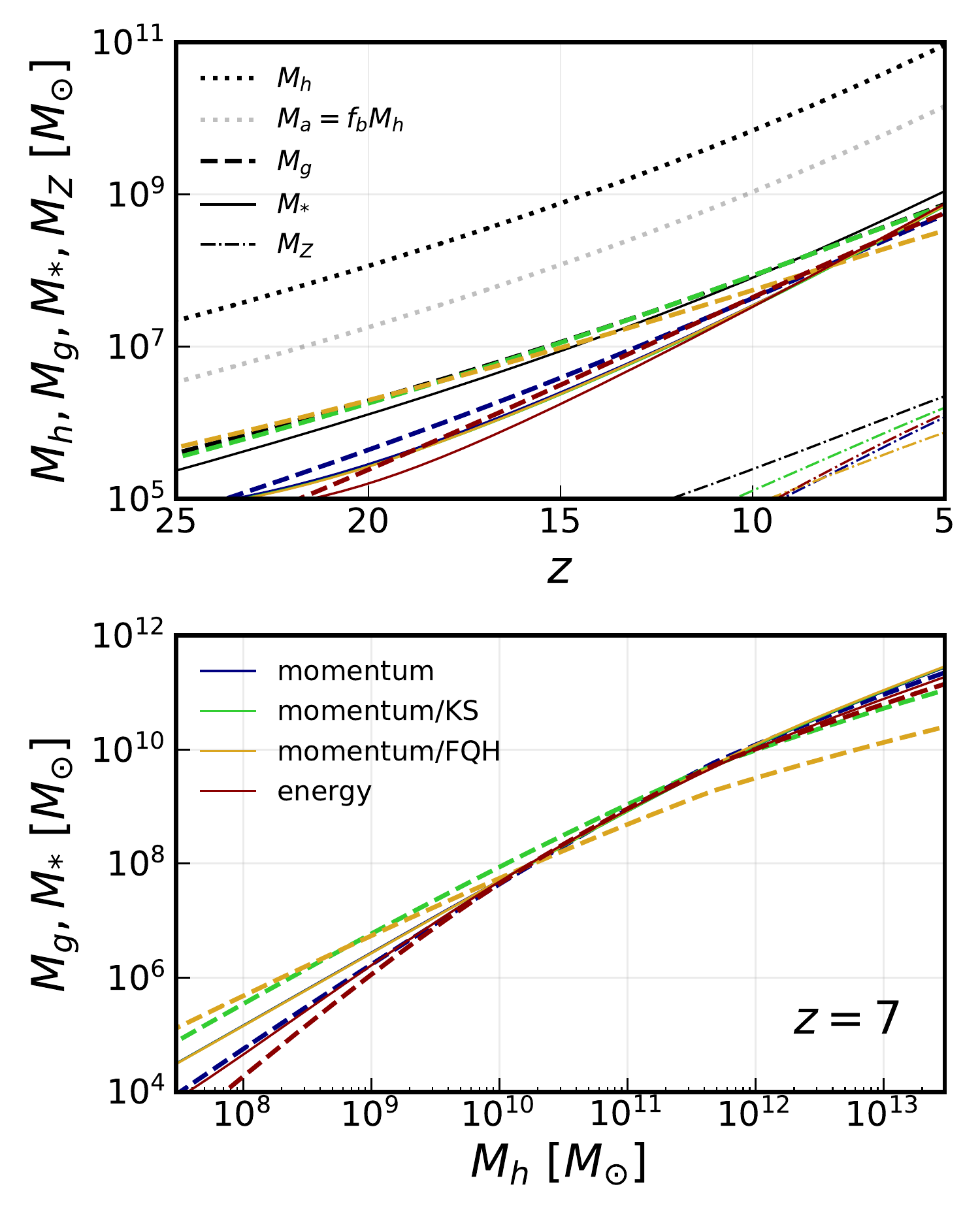}
 \caption{Top: the mass growth histories of a sample halo reaching $M \approx 10^{11}\,M_\odot$ at $z=5$ calculated by solving the system of differential equations from $z_i=30$. The black set of curves show halo properties calculated by the bathtub model with constant $\eta=10$ and $\epsilon=0.015$. The sets of curves in blue, green, yellow, and red show the results of models ``momentum'' (Model Ia), ``momentum/KS'' (Model Ib), ``momentum/FQH'' (Model Ic), and ``energy'' (Model II), respectively. For reference, the dotted curve in grey indicates the total accreted baryonic mass, which is more than 10 times larger than the mass of gas and stars formed as a result of strong feedback regulation. Bottom: the gas/stellar mass--halo mass relations at $z=7$ predicted by different choices of the feedback mode and star formation law.}
 \label{fig:halo_growth}
\end{figure}

\begin{figure*}
 \centering
 \includegraphics[width=0.98\textwidth]{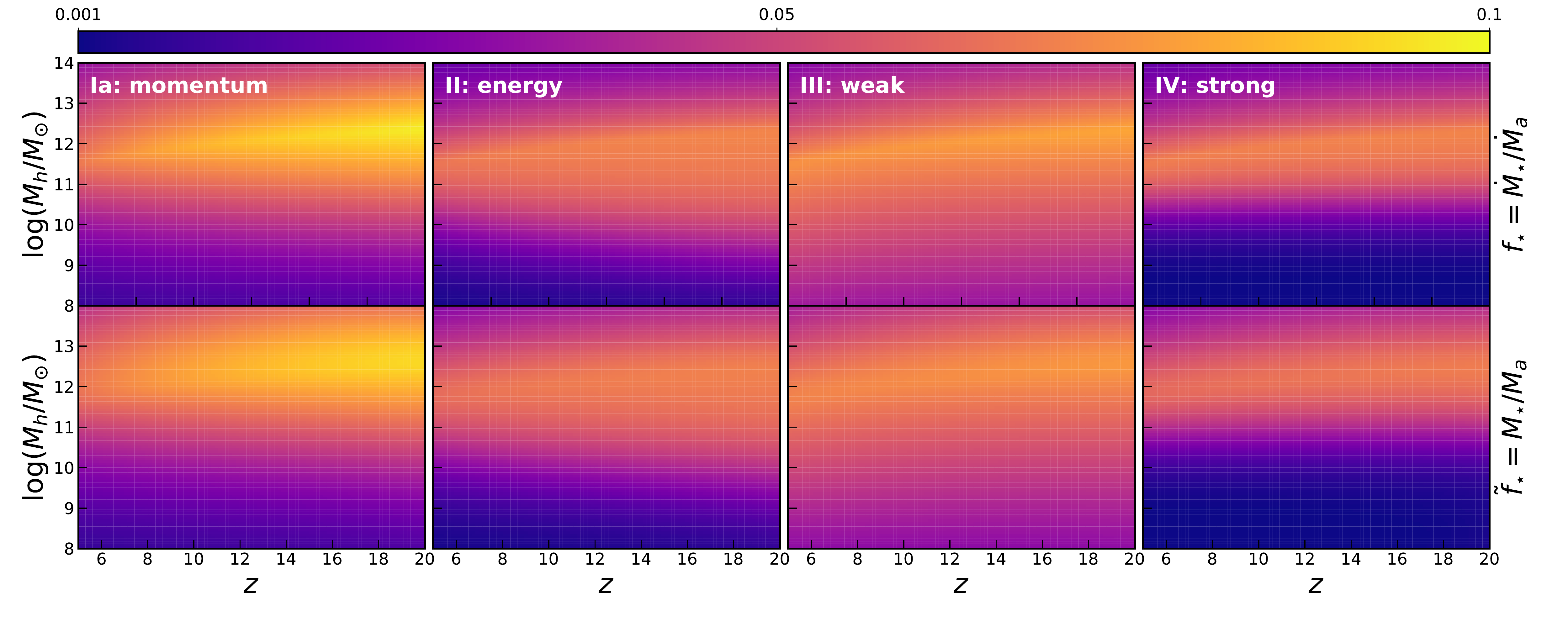}
 \caption{The mass and redshift dependence of the instantaneous (top row) and time-averaged (bottom row) star formation efficiencies calculated using different feedback prescriptions.}
 \label{fig:sfe}
\end{figure*}

To investigate what astrophysics of galaxy formation may be inferred from LIM data, we consider a total of 6 model variations involving different assumptions for the underlying feedback mode and star formation law, which are described by the value (or functional form) of the mass loading factor, $\eta$, and the temporal efficiency factor, $\epsilon$, respectively. Our baseline model assumes that stellar feedback is momentum-driven ($\eta \propto M^{-2/3}(1+z)^{-1}$) and sets $\epsilon$ to a fiducial value of 0.015 consistent with local observations \citep{KDM_2012}, which is referred to as Model~Ia and is the fiducial model assumed throughout Paper~I. A set of feedback variations are considered, where we explore a range of feedback modes leading to different star formation efficiency (SFE), especially in low-mass halos, while fixing $\epsilon$. In Model~II, we consider energy-driven feedback ($\eta \propto M^{-1/3}(1+z)^{-1/2}$), whereas in Model~III (IV) we envisage a more extreme scenario where the coupling between stellar feedback and the star-forming gas is weaker (stronger) than the momentum-driven (energy-driven) case. Specifically, a weak coupling in Model~III assumes $\eta \propto M^{1/6}$, whereas a strong coupling in Model~IV assumes $\eta \propto M$, with the redshift dependence being dropped in both cases for simplicity. 

\begin{table}
\centering
\caption{Specifications of the baseline model (Model~Ia, adopted for the fiducial simulations presented in Paper~I) and its variations considered in this work. The value of the escape fraction is varied accordingly to ensure that reionization completes by $z\approx5.5$. }
\label{tb:model_params}
\begin{tabular}{cccc}
\toprule
\toprule
Model & Feedback Mode & Star Formation Law & $f_{\rm esc}$ \\
\hline
Ia & momentum & $\epsilon=0.015$ & 10\% \\
Ib & momentum & KS & 10\% \\
Ic & momentum & FQH & 10\% \\
II & energy & $\epsilon=0.015$ & 12.5\% \\
III & weak & $\epsilon=0.015$ & 5\% \\
IV & strong & $\epsilon=0.015$ & 30\% \\
\bottomrule
\end{tabular}
\end{table}

On the other hand, a set of star formation law variations are explored in Models~Ib and Ic for momentum-driven feedback, where we further allow $\epsilon$ to vary moderately with the gas mass and thus yield star formation surface density--gas surface density relations\footnote{Throughout this work, we define star formation law to be the relation between the star formation rate surface density $(\dot{\Sigma}_{*})$ ans the gas surface density $(\Sigma_\mathrm{g})$, for which we follow \citet{Furlanetto_2021} to compute the galaxy size from the half-mass radius of the host halo, assuming that galaxy disc has uniform surface density.} corresponding to those implied by observations and/or theoretical predictions. In Model~Ib, we adopt $\epsilon \propto M_g^{0.2}$ which reproduces the well-known Kennicutt-Schmidt law \cite[][]{Kennicutt_1998} with a power-law index of 1.4, whereas in Model~Ic we adopt $\epsilon \propto M_g^{0.4}$ to approximate the star formation law with a power-law index of 2 as suggested by \citet{FQH_2013}, where the gas disc is assumed to be entirely turbulence-supported by stellar feedback. 

Solving Equations~(\ref{eq:mh_prime})--(\ref{eq:ms_prime}), we obtain the growth histories of stellar and gas masses in dark matter halos as they continuously accrete from an initial redshift of $z_i=30$. Figure~\ref{fig:halo_growth} shows the growth histories of gas, stellar, and metal masses for a sample dark matter halo with $M \approx 10^{11}\,M_\odot$ at $z=5$, derived from models with different feedback and star formation law combinations considered in this work. An averaged star formation efficiency (SFE) can be defined consequently as the stellar mass--halo mass ratio, namely $\tilde{f}_* = M_*/M_a$, 
which can be interpreted as the time-integrated value of the instantaneous SFE $f_* = \dot{M}_*/\dot{M}_a$ --- both are often derived in the literature with halo abundance matching (HAM), namely by matching the halo number density described by the halo mass function to the galaxy number density described by the galaxy UV luminosity function \cite[e.g.,][]{MTT_2015, Mashian_2016, SF_2016, Furlanetto_2017, Tacchella_2018, Behroozi_2019}. For simplicity, we limit the stellar population resulting from our star formation prescriptions to Pop~II stars only, and defer a systematic introduction of Population~III (Pop~III) stars to future studies (see Section~\ref{sec:discuss:extension}). Figure~\ref{fig:sfe} shows the instantaneous and time-averaged SFEs derived in different feedback prescriptions as a function of halo mass and redshift. The steepness of the color gradient indicates how strongly the fraction of accreted gas turned into stars is regulated by stellar feedback. 

\begin{figure*}[!ht]
 \centering
 \includegraphics[width=0.95\textwidth]{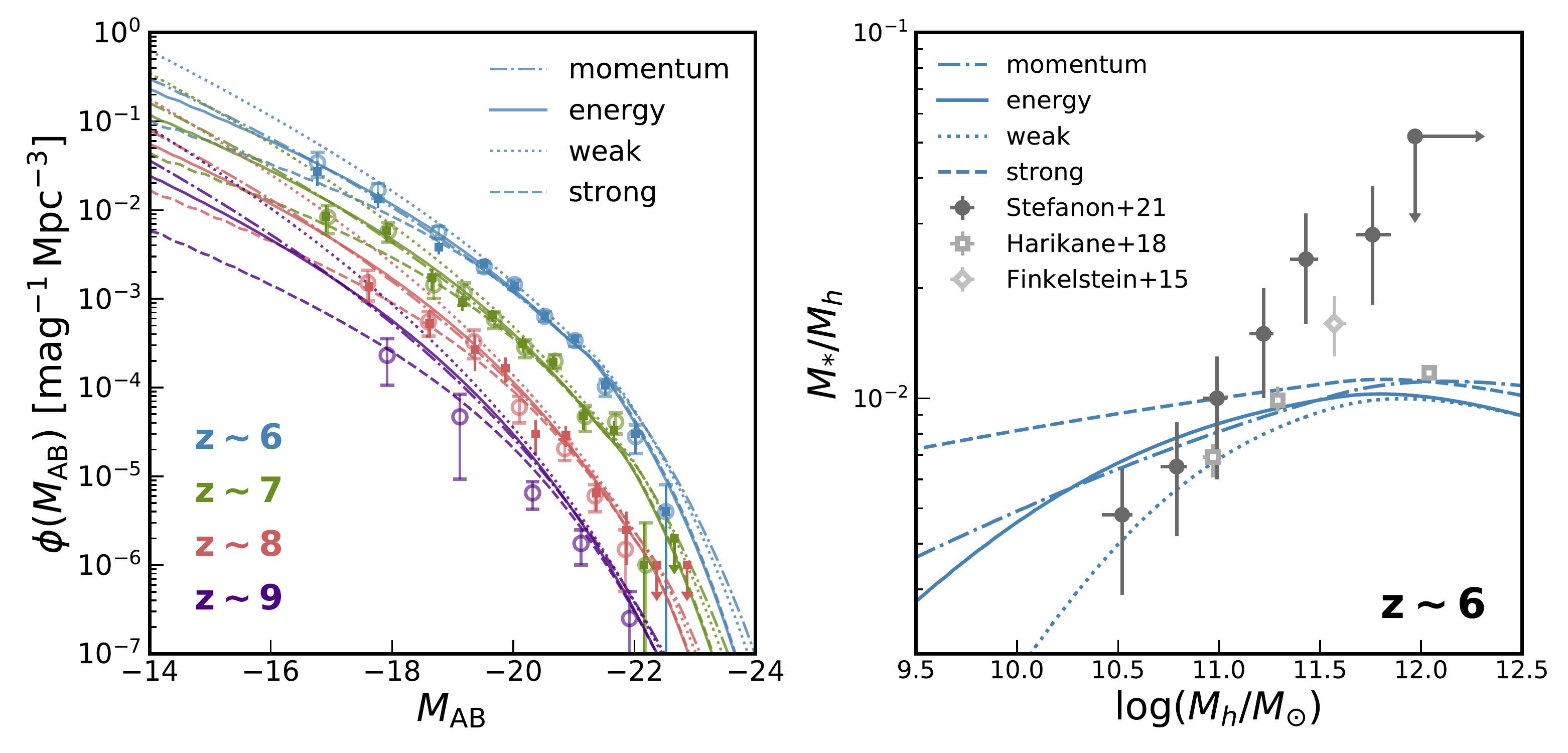}
 \caption{Left: galaxy rest-frame UV luminosity functions under different feedback assumptions. The predicted luminosity functions are compared against the observational constraints from \citet{Bouwens_2015} and \citet{Bouwens_2021}, as represented by the filled squares and empty circles, respectively. The $z=6$, $z=8$ and $z=9$ luminosity functions are offset vertically by a multiplicative factor of 2, 0.5 and 0.25, respectively, for ease of comparison. Right: comparison of the galaxy stellar-to-halo mass ratios implied by different feedback assumptions to the latest estimates from observations based on clustering analysis \citep{Harikane_2018} and HAM \citep{Finkelstein_2015, Stefanon_2021}.}
 \label{fig:uvlf}
\end{figure*}

In Figure~\ref{fig:uvlf}, we show a comparison between the observed galaxy UV luminosity functions and stellar-to-halo mass relation (SHMR) and our model predictions at $z \gtrsim 6$. As illustrated in the left panel, we verify that luminosity functions implied by the four feedback prescriptions considered are all reasonably well-consistent with constraints on the faint end from the Hubble Space Telescope (HST) data \citep{Bouwens_2015, Bouwens_2021}. In the right panel, we show that the SHMRs predicted by our model variations are roughly consistent with observations in the low-mass regime. At the high-mass end, our predictions only agree well with estimates based on the galaxy clustering \citep{Harikane_2018}, but not those based on HAM \citep{Finkelstein_2015, Stefanon_2021}, which are a factor of 2--3 larger. We note that lots of these uncertainties associated with EoR galaxies will be greatly reduced by new-generation telescopes like the James Webb Space Telescope (JWST), but as will be illustrated in what follows the information from multi-tracer LIM observations, which cover much wider areas wherein the entire galaxy population is accessed, will still be extremely valuable and complementary. 

\subsection{A Multi-Phase Extension of the Nebula Model}

As described in Paper~I, the numerical photoionization code \textsc{\textsc{cloudy}} \citep[version 17.02,][]{Ferland_2017} is supplemented to galaxy properties predicted by the galaxy formation model in LIMFAST to simulate the production of various emission lines as target LIM signals. Here, we extend the baseline nebular model introduced in Paper~I, which mainly accounts for lines produced in \ion{H}{2} regions, to include bright emission lines of particular interest to LIM studies from the neutral (atomic/molecular) ISM, such as [\ion{C}{2}] 158\,$\mu$m, and CO(1--0) 2601\,$\mu$m lines. We note that because any legitimate nebular model based on \textsc{\textsc{cloudy}} can be used as the input of LIMFAST, in what follows we do not repeat the analysis of optical/UV lines discussed in Paper~I with the new nebular model. For H$\alpha$, Ly$\alpha$, and [\ion{O}{3}] 88\,$\mu$m lines considered in this work, we simply reuse the results from Paper~I, although in principle they can be captured together with lines originating from the neutral ISM by a generalized nebular model.  

\begin{figure}
 \centering
 \includegraphics[width=0.48\textwidth]{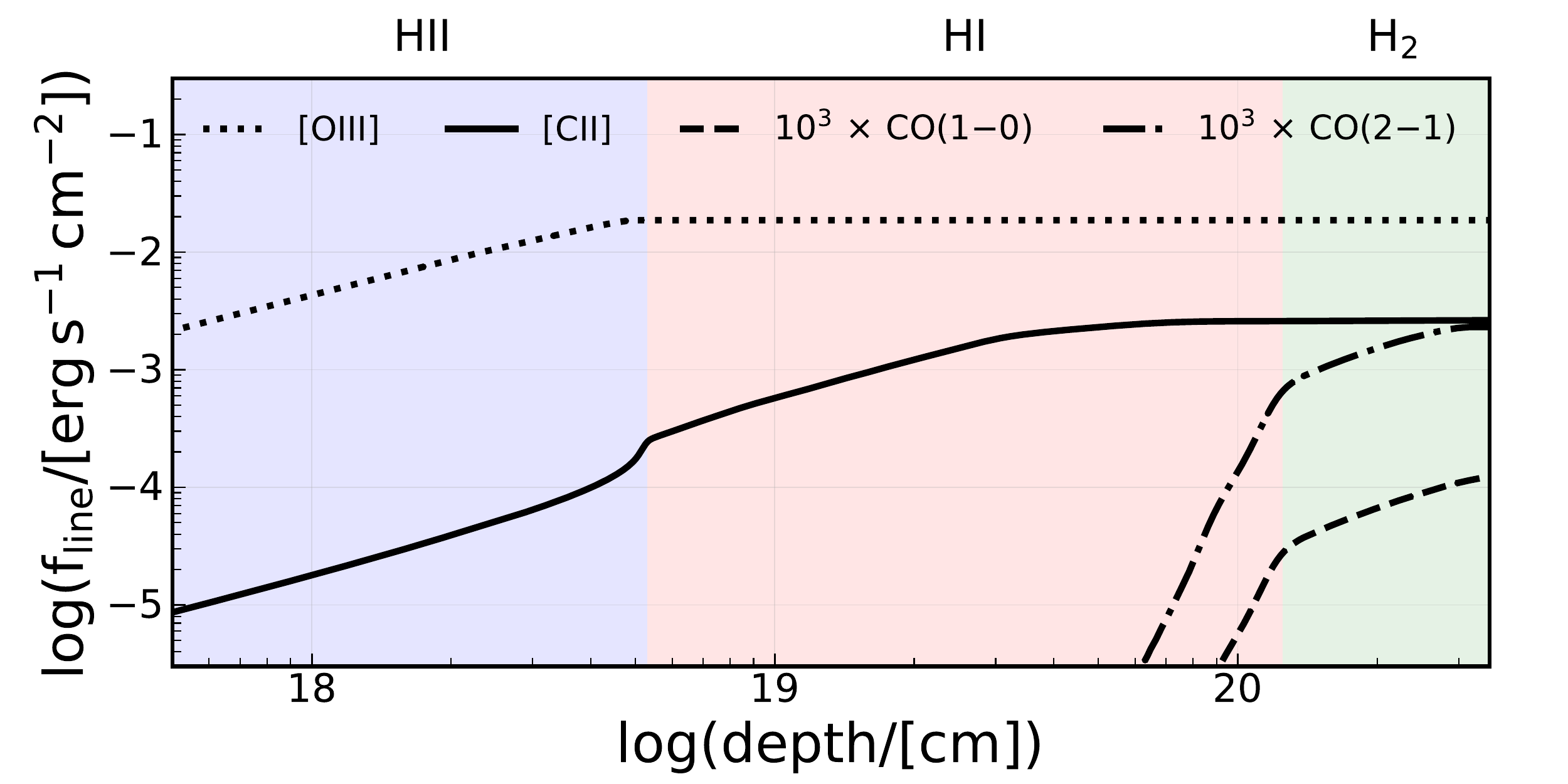}
 \caption{Cumulative emissivities of [\ion{O}{3}] 88\,$\mu$m, [\ion{C}{2}] 158\,$\mu$m, CO(1--0) 2601\,$\mu$m, and CO(2--1) 1300\,$\mu$m lines calculated with \textsc{\textsc{cloudy}} assuming a gas cloud with density $n_\mathrm{H}=100\,\mathrm{cm^3}$ and metallicity $Z = 0.002$ illuminated by the interstellar radiation field $G_0=10^3$ in  the Habing flux. Note that CO lines are enlarged by 1000 times for the ease of comparison. Background colors indicate different ISM phases where the lines predominantly originate from, with the boundaries corresponding to where sharp changes in the gas kinetic temperature profile occur. }
 \label{fig:cloudy_emissivity}
\end{figure}

For lines emitted from atomic or molecular gas, namely in PDRs or $\mathrm{H}_2$ cores of gas clouds, because their strengths depend on the gas content, we define an equivalent surface area $\mathcal{S}$ according to the distribution of gas density in giant molecular clouds (GMCs) following the prescription from \citet{Vallini_2018}. For simplicity, the gas mass $M_g$ of a given halo is assumed to be evenly distributed among the population of GMCs with the same fixed mass $M_{\rm GMC}$, such that the total line luminosity can be simply scaled from that of one single GMC. Specifically, to describe the internal structure of GMCs, we first define a volumetric distribution of gas density $\rho$ that follows a log-normal probability distribution function (PDF), as suggested by models of isothermal, non-self-gravitating, turbulent gas \cite[e.g.,][]{PV_1998, PN_2002}. Namely, 
\begin{equation}
P_V(\rho) \propto \frac{1}{\sqrt{2\pi}\sigma} \exp{\left[ - \frac{\left(\ln(\rho/\rho_0)-\langle \ln(\rho/\rho_0) \rangle\right)^2}{2\sigma^2} \right]}~,
\end{equation}
where $\rho_0 = \mu m_{\rm p} n_{\rm H,0}$ is the mean gas density of the GMC, with $n_{\rm H,0} = 100\,\mathrm{cm^{-3}}$ and $\mu=1.36$ accounting for helium. The logarithmic scatter $\sigma$ satisfies $\langle \ln(\rho/\rho_0) \rangle = -\sigma^2/2$, maintaining a fixed expectation of $\ln({\rho/\rho_0})$ as $\sigma$ varies. The distribution of $\sigma$ depends on the turbulence level characterized by the Mach number $\mathcal{M}$ through
\begin{equation}
\sigma^2 = \ln(1+b^2 \mathcal{M}^2)~,
\end{equation}
where $b=0.3$ describes the efficiency of turbulence production and we take $\mathcal{M}=5$, a plausible value for high-redshift galaxies that tend to have more supersonic structures \cite[see discussion in e.g.,][]{SS_2016}. Since in this work we focus on lines with low to intermediate critical densities (rather than e.g., high-$J$ CO lines tracing the densest regions in GMCs), we ignore self-gravity, which has the critical effect of modifying the density distribution into a power law at high enough densities. The PDF is normalized such that its integral over gas density gives the total volume of the GMC, 
\begin{equation}
V_{\rm GMC}^{\rm tot} = \int d V = \int P_V(\rho|\mathcal{M}) d \rho = \frac{4\pi}{3}R^3_{\rm GMC}~,
\end{equation}
where $R_{\rm GMC} = 15\,$pc specifies the size of one GMC, implying a mass of $M_{\rm GMC} = \rho_0 V_{\rm GMC}^{\rm tot} = 4.7\times10^4\,M_\odot$. This allows one to conveniently define a characteristic length scale corresponding to each density $\rho$, 
\begin{equation}
r(\rho) = \left[ \int_{\rho - \delta \rho}^{\rho + \delta \rho} P_V(\rho'|\mathcal{M}) d \rho' \right]^{1/3}~,
\end{equation}
with which an equivalent surface area $\mathcal{S}=4\pi r^2$ can be defined for the cumulative line flux $f_{\rm line} = 4\pi j_{\rm line}$ in units of $\mathrm{erg\,s^{-1}\,cm^{-2}}$. The line luminosity density (i.e., $l_{\rm line} \equiv d L_{\rm line}/d V$) can be then expressed as
\begin{equation}
l_{\rm line}(\rho) = \frac{4\pi r^2(\rho)}{V_{\rm GMC}^{\rm tot}} f_{\rm line}(\rho, Z_g, U)~,
\end{equation}
where $f_{\rm line}(\rho, Z_g, U)$ means that the line luminosity is a function of gas density, gas metallicity, and ionization parameter in \textsc{\textsc{cloudy}}. Figure~\ref{fig:cloudy_emissivity} shows the cumulative emissivities of the three FIR/mm-wave lines considered in this work, for a gas density of $n_\mathrm{H,0}=100\,\mathrm{cm^3}$ and an interstellar radiation field (ISRF) of strength $G_0 = 10^3$ in units of the Habing flux, which we adopt as fiducial parameters in our model, consistent with observations of the ISM at high redshifts \citep{Gullberg_2015, Wardlow_2017}. The line luminosity of the GMC is consequently given by
\begin{equation}
\mathcal{L}^{\rm tot}_{\rm line} = \int l_{\rm line}(\rho) P_V(\rho|\mathcal{M})\,d \rho~.
\end{equation}
Finally, to arrive at the total line luminosity of a halo of gas mass $M_g$, we simply scale $\mathcal{L}^{\rm tot}_{\rm line}$ by
\begin{equation}
L^{\rm tot}_{\rm line} = \mathcal{L}^{\rm tot}_{\rm line} M_g / M_{\rm GMC}~.
\end{equation}

\begin{figure*}
 \centering
 \includegraphics[width=\textwidth]{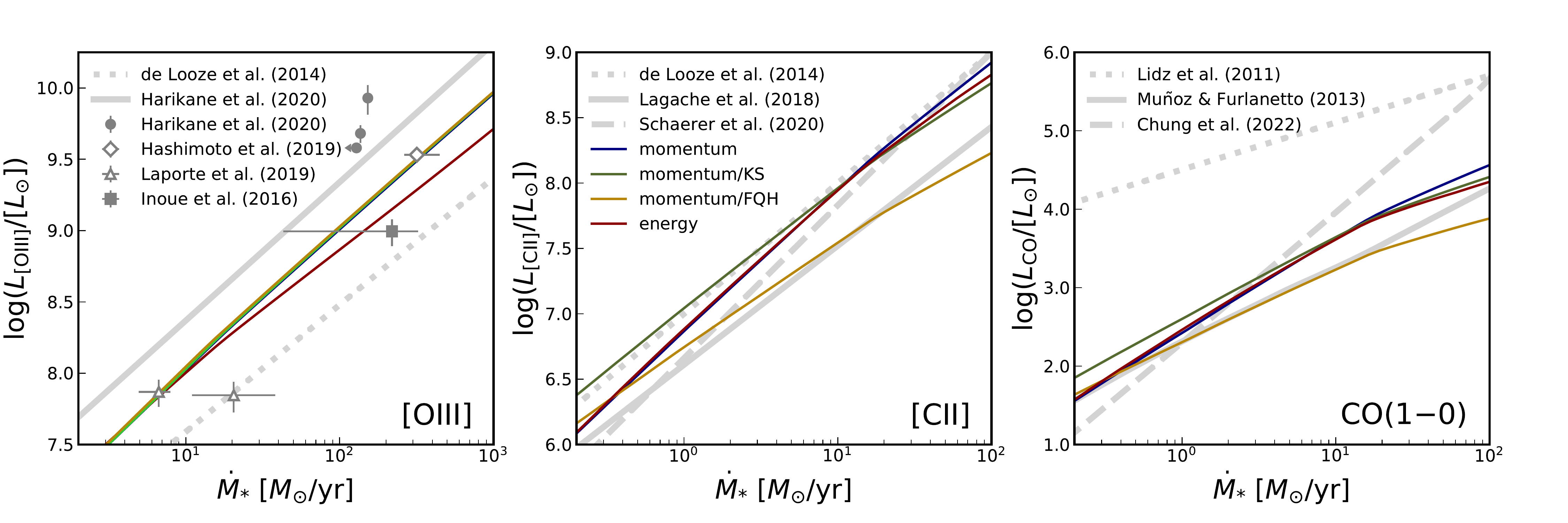}
 \caption{The luminosity--star formation rate relations of [\ion{O}{3}], [\ion{C}{2}], and CO(1--0) lines at $z\sim6$ predicted by LIMFAST assuming different stellar feedback and star formation law prescriptions. Empirical fittings to observations of low-/intermediate-$z$ galaxies \cite[][]{Lidz_2011, deLooze_2014, Chung_2022} and high-$z$ galaxies \cite[][]{Harikane_2020, Schaerer_2020}, as well as predictions from physically-motivated high-$z$ ISM models \cite[][]{Lagache2018, MF_2013} from the literature are shown by the gray lines. Because of the relatively dearth of high-$z$ [\ion{O}{3}] emitters observed, a selected number of recent [\ion{O}{3}] observations at $z>6$ are also shown for comparison \citep{Inoue_2016, Hashimoto_2019, Laporte_2019, Harikane_2020}.}
 \label{fig:lsfr}
\end{figure*}

Figure~\ref{fig:lsfr} shows the luminosity--star formation rate relations for [\ion{O}{3}], [\ion{C}{2}], and CO lines from $z=6$ (solid) and $z=8$ (dashed) galaxies as predicted by our galaxy models. For comparison, we also plot empirical representations of the observed/inferred luminosity--star formation rate relations for low-/intermediate-$z$ galaxies \cite[][]{Lidz_2011, deLooze_2014,Chung_2022}, together with high-$z$ fitting relation \citep{Harikane_2020} and predictions based on physically-motivated ISM models \cite[][]{MF_2013, Lagache2018}. For [\ion{C}{2}] emission, recent observations of $4 \la z \la 6$ galaxies \cite[e.g.,][]{Schaerer_2020} suggest a lack of evolution in the $L_\mathrm{[CII]}$--SFR relation from that in the local universe \cite[][]{deLooze_2014}, which slightly disfavors some theoretical predictions derived from a combination of semi-analytical models of galaxies and photoionization simulations \citep{Lagache2018}. Notably for CO(1--0) emission, the models displayed from LIMFAST and the literature differ significantly in terms of the shape and normalization, partly because of the different redshift regimes where these models are evaluated/calibrated. Future CO galaxy and LIM surveys with ngVLA and COMAP \cite[see e.g.,][and references therein]{Breysse_2022} will greatly improve the constraints on scaling relations like the one shown here.

\subsection{Emission Lines From the IGM} \label{sec:model:igm}

Besides nebular emission lines produced in the multi-phase ISM, for the IGM emission LIMFAST also inherits and improves on the detailed 21 cm calculations from 21cmFAST, while adopting a simple prescription for recombination emission from the diffuse, ionized IGM in Ly$\alpha$. As detailed in Sections 2.5 and 2.6 of Paper~I, where interested readers are referred to for a complete description, the modeling of these emission lines from the IGM is also fully coupled with the galaxy formation model and its variations implemented in LIMFAST. This allows the influence of galaxy astrophysics on the statistics of 21 cm and IGM Ly$\alpha$ emission to be studied self-consistently with other emission lines from the ISM as tracers of the galaxy distribution.

\begin{figure*}[!ht]
 \centering
 \includegraphics[width=0.95\textwidth]{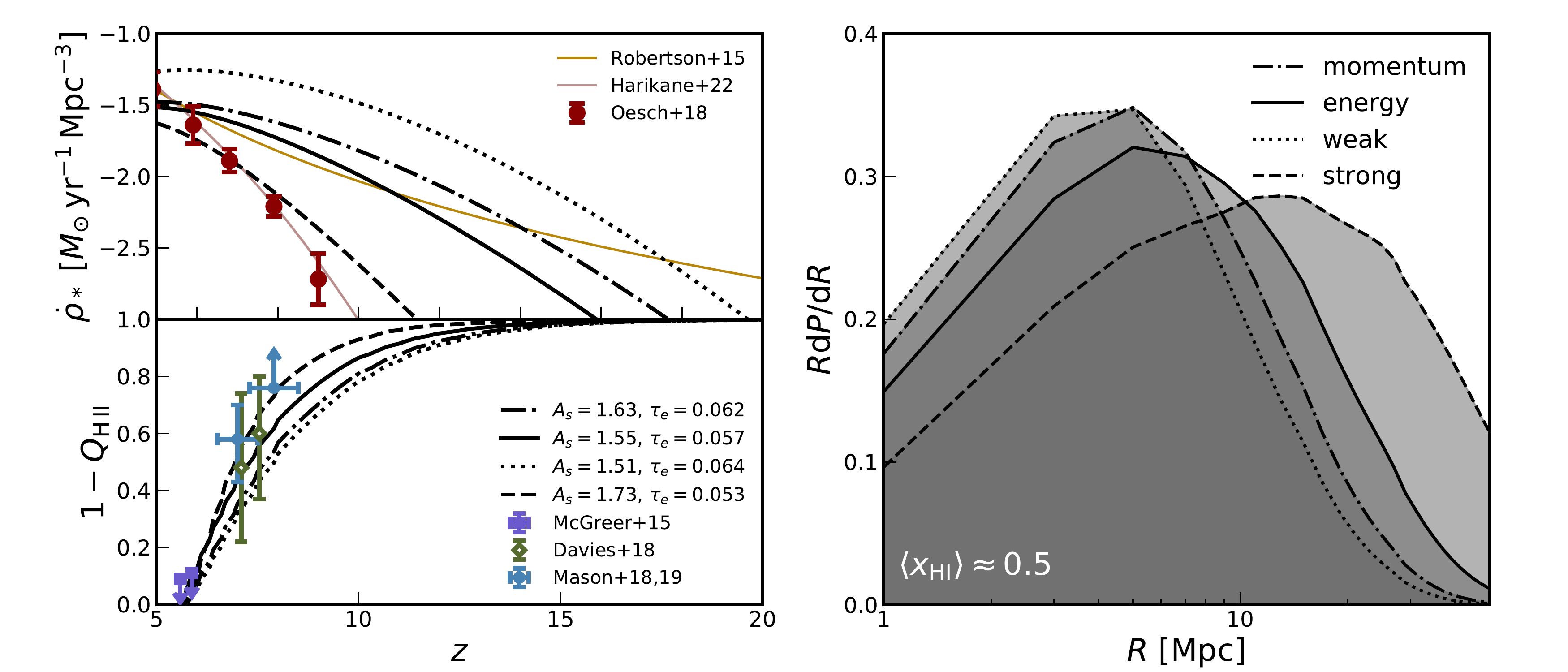}
 \caption{Left: the redshift evolution of the cosmic SFRD (top) and the mean neutrality of the IGM (bottom) simulated by LIMFAST models with different feedback assumptions. Also shown in the top panel are observational constraints on the SFRD from \citet{Oesch_2018} integrated down to $M_{\rm UV}=-17$, along with empirical fits from \citet{Robertson_2015} and \citet{Harikane_2022} assuming different amounts of extrapolation for the faint populations. Observational constraints on the mean IGM neutral fraction from the dark fraction in the Ly$\alpha$ and Ly$\beta$ forests \citep{McGreer_2015}, IGM damping wing signatures in quasar spectra \citep{Davies_2018}, and Ly$\alpha$ emission from Lyman-break galaxies \citep{Mason_2018, Mason_2019} are shown in the bottom panel. Curves of different feedback mode are also labeled by an asymmetry measure, $A_\mathrm{s}$, of the EoR history, and the implied CMB electron scattering optical depth, $\tau_{e}$, which is verified to be consistent with the latest observational constraints \citep{Pagano_2020, Qin_2020b}. Right: the bubble size distribution derived with the ``mean free path'' approach when the IGM is approximately half-ionized.}
 \label{fig:eor}
\end{figure*}

\section{Results} \label{sec:results}

In this section, we present the main quantitative results of this paper derived from the set of model variations specified in Table~\ref{tb:model_params}. We first show the global reionization histories implied by models with varying feedback prescriptions (Section~\ref{sec:results:eor}), which supplements the model predictions presented in Paper~I based on Model~Ia only, and then present how the corresponding sky-averaged signals of various lines are sensitive to the changes in feedback. Next, we demonstrate how variations of the feedback mode (Section~\ref{sec:results:feedback}) and the star formation law (Section~\ref{sec:results:sflaw}) in play affect summary statistics, namely the auto- and cross-correlation power spectra of tracers of neutral and ionized IGM. By examining the shape and amplitude evolution of power spectra, we elaborate on how astrophysical information about ionizing sources and the IGM may be extracted in turn from joint analyses of multi-tracer LIM observations. For clarity, all results presented in the remainder of this section are shown in real space, without considering observational effects such as redshift space distortions (RSDs), whose treatment in LIMFAST is elaborated in Section~2.7 of Paper~I.

\subsection{Reionization Histories} \label{sec:results:eor}

\begin{figure*}
 \centering
 \includegraphics[width=\textwidth]{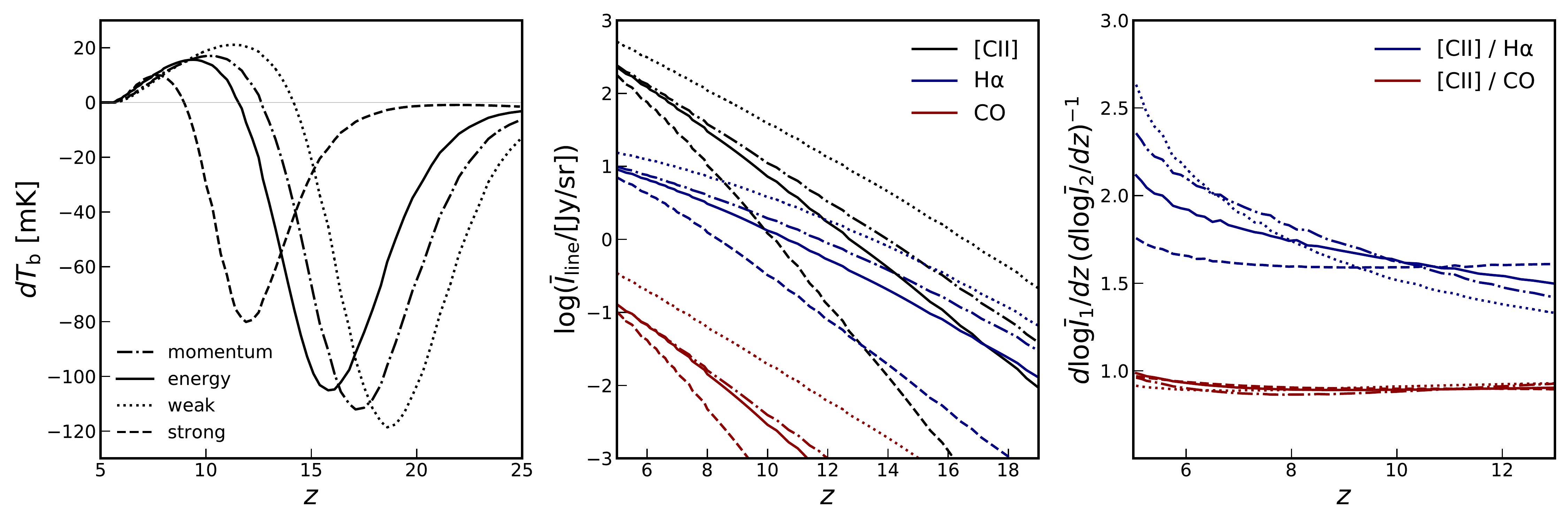}
 \caption{The redshift evolution of the sky-averaged (global) 21 cm different brightness temperature (left) and intensities of [\ion{C}{2}], H$\alpha$ and CO lines (middle) under different feedback assumptions. Also shown in the right panel is the ratio of redshift dependence of different line intensities, which serves as a measure of the feedback-sensitive metal enrichment history.}
 \label{fig:skyavg}
\end{figure*}

Effects of feedback regulation on star-forming galaxies as ionizing sources can be revealed by both the global history and the detailed morphology of reionization. Figure~\ref{fig:eor} shows two important measures of the reionization, the volume-averaged neutral fraction and the ionized bubble size distribution (BSD), simulated by LIMFAST assuming different feedback prescriptions, which yield the different redshift evolution of the cosmic SFRDs shown in the upper left panel. Notably, the cosmic SFRD directly relates to the strength of stellar feedback through the SFE in low-mass halos. As illustrated in Figure~\ref{fig:sfe}, more efficient feedback coupling results in a steeper SFE gradient with halo mass, implying less efficient star formation in low-mass halos and thus an overall lower and steeper cosmic SFRD dominated by massive halos. The SFRDs predicted by our momentum- and energy-regulated models are comparable to the extrapolation to observations from \citet{Robertson_2015} out to $z\sim15$ (despite the opposite curvature), whereas stronger or weaker feedback can result in SFRDs close to or substantially higher ($>1$\,dex) than the observational constraints available to date \citep{Oesch_2018, Harikane_2022}. Since we tune $f_\mathrm{esc}$ such that the reionization completes roughly at the same time at $z\approx6$ in each feedback scenario (see Table~\ref{tb:model_params}), a steeper SFRD corresponds to an overall more rapid and asymmetric reionization history, as measured by the factor $A_\mathrm{s} \equiv (z_{05}-z_{50})/(z_{50}-z_{95})$ which uses the reionization midpoint $z_{50}$ and 5\% (95\%) completion point $z_{05}$ ($z_{95}$) to characterize the asymmetry of the full extent of the EoR \citep{Trac_2018}. 

The impact of stellar feedback on the size of ionized regions is illustrated in the right panel of Figure~\ref{fig:eor}, which shows the BSD in different feedback modes when the IGM is about half-ionized. Following the ``mean free path'' method introduced by \citet{MF_2007}, we describe the BSD with the probability density function of the logarithmic bubble radius $R$, which is calculated by repeatedly sampling the size of \ion{H}{2} regions from random ionized points and in random directions with a Monte Carlo process. At a fixed neutral fraction $\langle x_{\rm H \textsc{i}} \rangle$, the BSD shifts towards larger bubble radius when the feedback regulation is stronger. Although a mass-dependent $f_\mathrm{esc}$ may lead to a similar effect, the degeneracy can be reduced by direct constraints on $f_\mathrm{esc}$ from future observations of individual EoR galaxies by e.g., JWST.  As will be shown in what follows, cross-correlations between the 21 cm signal and tracers of star-forming galaxies turn out to be sensitive probes of the typical bubble size encoded by the BSD, even though the exact correspondence relies on a good understanding of the astrophysics. 

\subsection{Sky-Averaged Line Intensities} \label{sec:results:skyavg}

The sky-averaged intensity of spectral line emission, especially that of the 21 cm line (often referred to as the global 21 cm signal), as a spatial monopole measurement is known to be a useful probe of the EoR history and source population \citep{Mirocha_2015, Mirocha_2017, Cohen_2017, MF_2019, Park_2019}. Figure~\ref{fig:skyavg} shows the redshift evolution of the sky-averaged signals of various lines and their ratios. For the 21 cm global signal, $\delta \bar{T}_\mathrm{b}(z)$, as revealed by the timing and strength of its extrema, a stronger feedback implies delayed Ly$\alpha$ coupling and heating, which lead to an overall smaller signal amplitude in both absorption ($\delta \bar{T}_\mathrm{b} < 0$) at cosmic dawn and emission ($\delta \bar{T}_\mathrm{b} > 0$) during cosmic reionization. The absorption trough varies between $z\sim12$ and 18 in the central redshift and between $\delta \bar{T}_\mathrm{b}=-80$ and $-120$\,mK in the depth for the four feedback modes considered, suggesting an intimate connection between feedback-regulated star formation in the first galaxies and the 21 cm spin temperature evolution during cosmic dawn. The is consistent with the overall shift of the 21 cm global signal towards lower redshift/higher frequency, as projected by the recent literature taking into account of the observed UV luminosity function at $z \gtrsim 6$ (\citealt{Mirocha_2017}; \citealt{Park_2019}; but see also \citealt{Cohen_2017}). 

\begin{figure*}[!ht]
 \centering
 \includegraphics[width=0.95\textwidth]{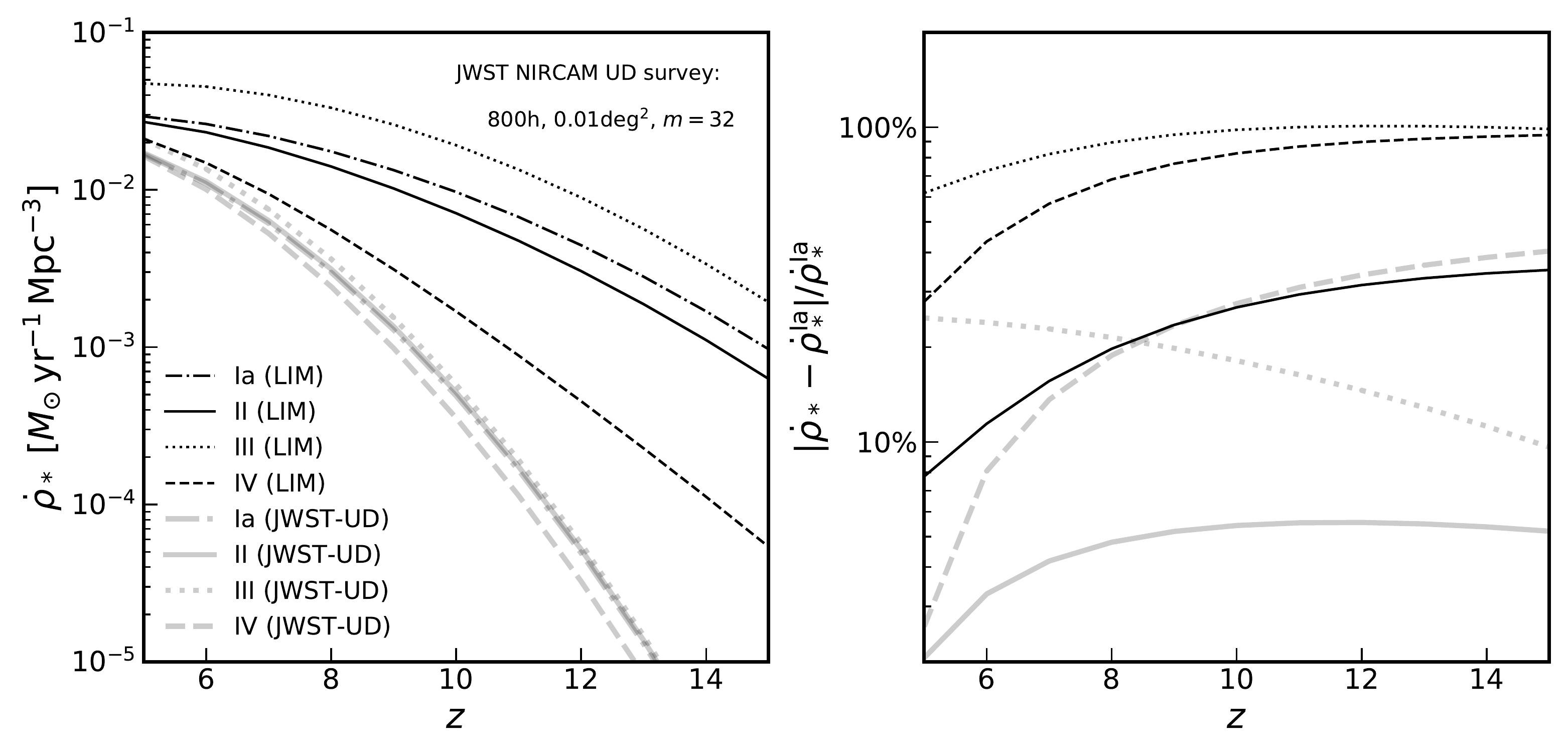}
 \caption{Left: a comparison between the cosmic SFRD attributed to galaxies measurable by LIM experiments and a nominal JWST/NIRCAM ultra-deep survey reaching a limiting magnitude of $m \approx 32$ with four 200\,h pointings covering a total area of $\approx 0.01\,\mathrm{deg^2}$. The thick and thin sets of curves represent LIM- and JWST-detectable galaxies, respectively. Right: the fractional SFRD deviation from the fiducial momentum-driven feedback model (Model~Ia) in other feedback models.}
 \label{fig:jwst}
\end{figure*}

The middle panel of Figure~\ref{fig:skyavg} shows the mean intensities of line tracers of galaxies [\ion{C}{2}], H$\alpha$, and CO, whose redshift evolution is largely driven by that of the SFRD. Nonetheless, the subtle difference in the steepness of redshift evolution, caused by the different metallicity dependence of these tracers, serves as a potential probe of feedback through the implicit metal enrichment history. The effect is illustrated in the right panel of Figure~\ref{fig:skyavg}, where ratios of slopes $\bar{I}^{-1} d \bar{I} / dz$ as a function of redshift are contrasted with each other. Clearly, the slope ratio of lines with high contrast in metallicity dependence (e.g., [\ion{C}{2}] and H$\alpha$) is sensitive to feedback, with less efficient feedback producing a slope ratio with stronger redshift evolution, whereas the slope ratio of lines with similar metallicity dependence (e.g., [\ion{C}{2}] and CO) is largely a constant insensitive to the exact feedback mechanism. Such sensitivity to feedback is not surprising though, given that the mean intensity evolution is mainly driven by the much more abundant low-mass galaxies that are most feedback-sensitive and least metal-enriched. In the case of very strong feedback, there are too few low-mass galaxies to make a significant variation in the slope ratio, and thus it remains roughly constant with redshift. 

Measurements of the mean intensity evolution of spectral line tracers of galaxies, especially SFR tracers like H$\alpha$ and [\ion{C}{2}], can often be translated into constraints on the SFRD, provided that the $L$--SFR relation can be reliably determined. This in turn provides an angle to compare the information from LIM observations to what will be available from forthcoming surveys of individual high-$z$ sources by new-generation telescopes like the JWST. In Figure~\ref{fig:jwst}, we illustrate a simple comparison between the distinguishing power on the cosmic SFRD in different feedback modes available from JWST and LIM observations in general. As an example, we consider an potential ultra-deep (UD) configuration for a galaxy dropout survey with JWST/NIRCAM that reaches a limiting magnitude of $m\approx32$, similar to the strategies considered in \citet{MTT_2015} and \citet{Furlanetto_2017}\footnote{Even though more realistic survey plans are now available \cite[see e.g.,][]{Williams_2018, Robertson2021arXiv}, the approximate configuration, as presented, is sufficient for our purpose.}. As shown in the left panel of Figure~\ref{fig:jwst}, a JWST UD survey will likely still miss a considerable fraction ($\gtrsim50\%$) of the total star formation in galaxies at $z\gtrsim6$, unless the SFRD declines steeply with redshift due to a shallow faint-end slope of the galaxy luminosity function, a likely result of very efficient stellar feedback. On the contrary, the statistical nature of LIM makes the measurements sensitive to the collective star formation activity sourcing the aggregate line emission, although in some cases the conversion between line luminosity and the SFR can be sophisticated.

\begin{figure*}[ht]
 \centering
 \includegraphics[width=0.8\textwidth]{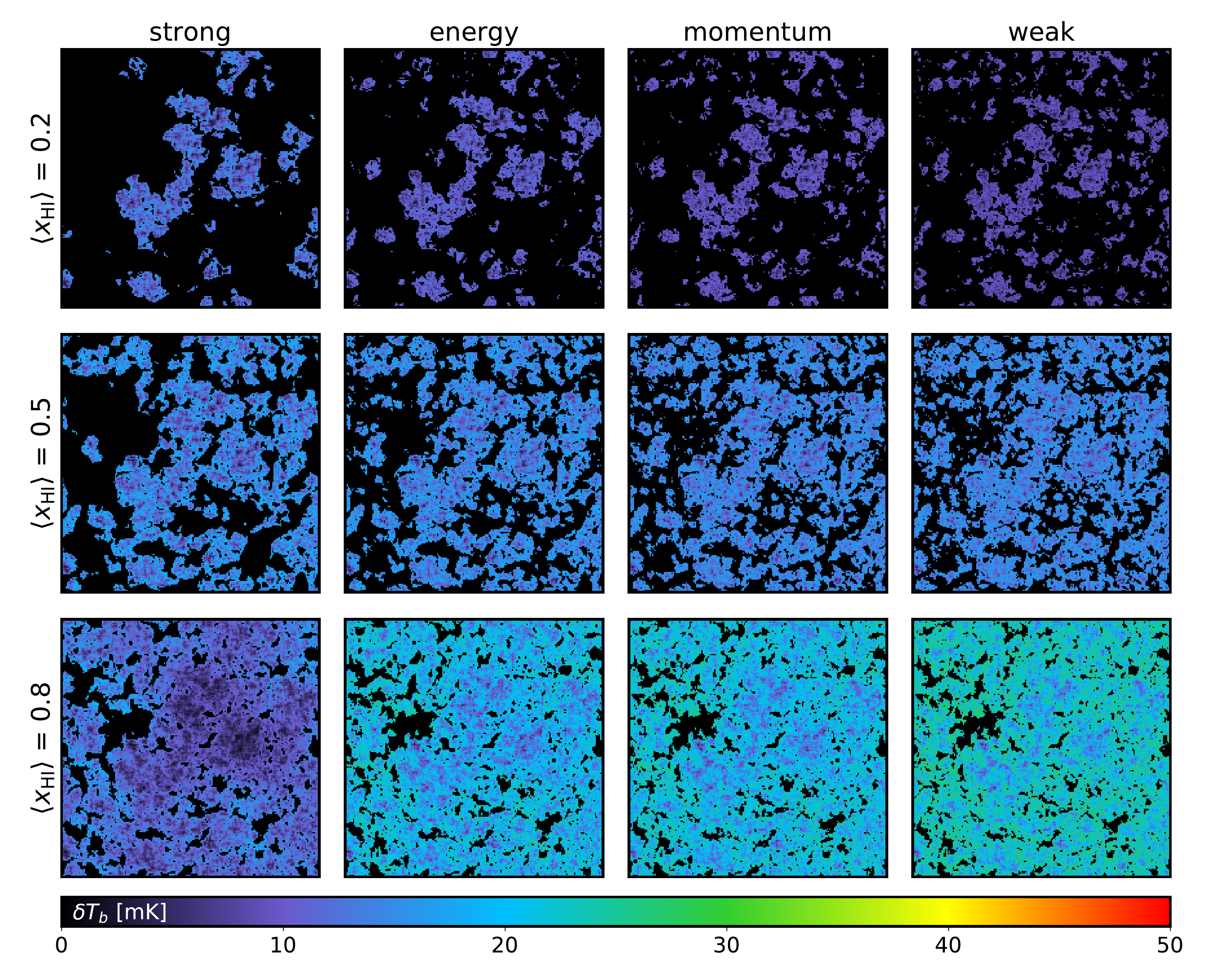}
 \includegraphics[width=0.8\textwidth]{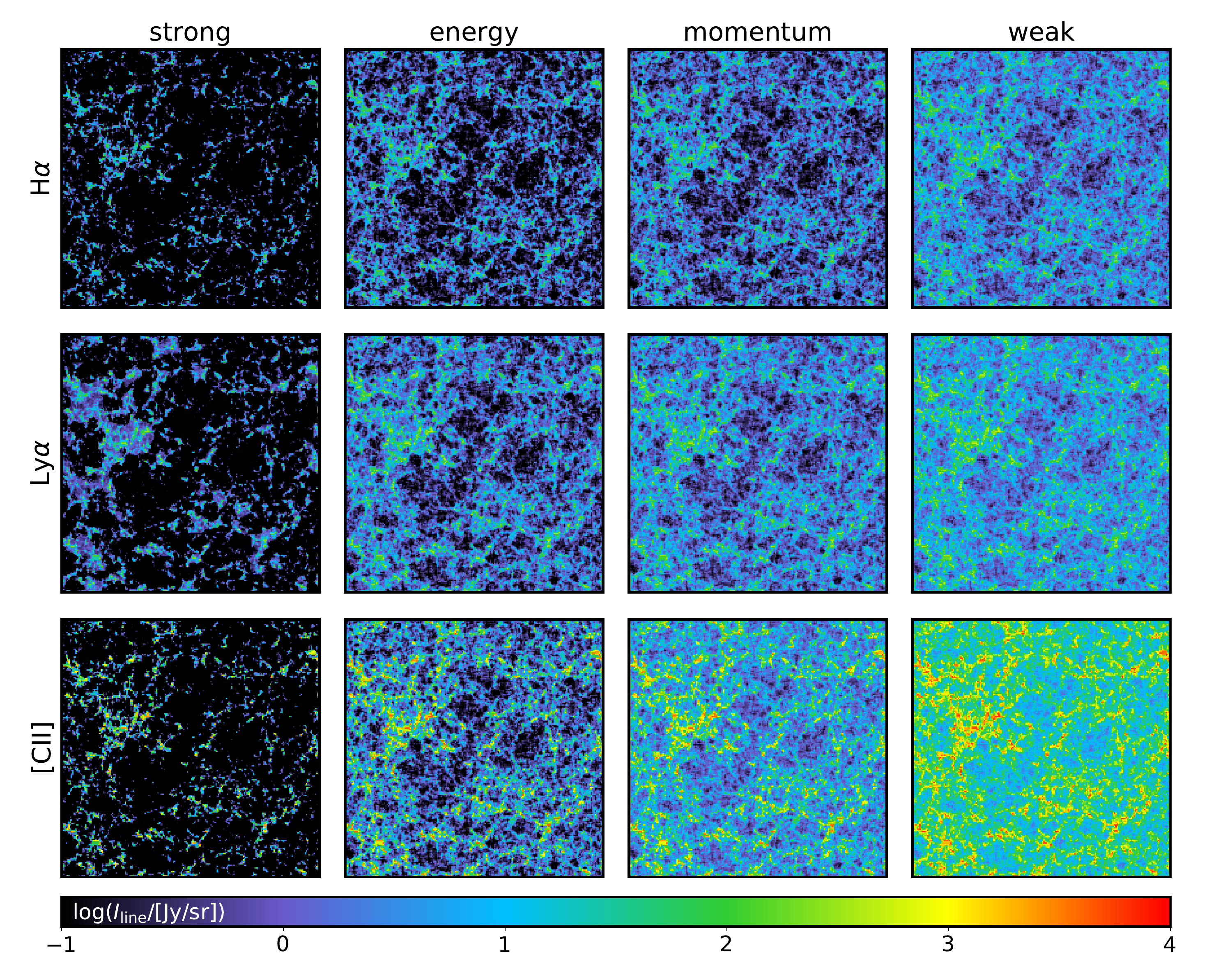}
 \caption{Left: slices of 21 cm differential brightness temperature $\delta T_b$ fields at various stages of reionization ($\langle x_{\rm H \textsc{i}} \rangle \approx 0.2$, 0.5, and 0.8, respectively) simulated by LIMFAST, assuming different stellar feedback prescriptions as specified in Table~\ref{tb:model_params}. Right: slices of H$\alpha$, Ly$\alpha$, and [\ion{C}{2}] line intensity fields at $z\approx7$ ($\langle x_{\rm H \textsc{i}} \rangle \approx 0.5$) simulated by LIMFAST. Note that the Ly$\alpha$ intensity fields displayed have contributions from both Ly$\alpha$ emitters and recombinations in the diffuse ionized IGM, though without accounting for the damping wing absorption due to the intervening neutral IGM. Each slice is 256\,Mpc on a side and 1\,Mpc thick.}
 \label{fig:snapshots}
\end{figure*}

To further contrast the two types of measurements in the context of probing stellar feedback, we show in the right panel of Figure~\ref{fig:jwst} the fractional deviation of the total, \textit{measurable} SFRD in other feedback modes from that in the fiducial, momentum-driven case of feedback. Namely, $|\dot{\rho}_*-\dot{\rho}^\mathrm{Ia}_*|/\dot{\rho}^\mathrm{Ia}_*$ reflects how easily one might disprove a simple momentum-driven feedback model using deviations (if any) of the observed SFRD from the expected one. Due to the insufficient sensitivity of galaxy surveys to faint objects, for which the effect of feedback regulation is most pronounced, a JWST UD survey tends to have less distinguishing power than LIM observations especially towards higher redshift. The exception, again, is when comparing a very strong feedback to Model~Ia, in which case the difference in distinguishing power decreases with increasing redshift as galaxies to which LIM is uniquely sensitive diminish rapidly, although LIM observations still offer more distinguishing power. We note that the example presented here only represents an extremely-simplified, special case of inferring stellar feedback from the SFRD evolution. In practice, the individual source detection and LIM methods further complement each other by the different quantities that are directly probed (e.g., the luminosity function vs. moments of the luminosity function) and the different sources of uncertainty involved (e.g., cosmic variance and the dust correction vs. foreground contamination and the $L$--SFR relation), and thus are both valuable probes of galaxy formation and evolution in the high-$z$ universe.

\subsection{Characterizing Stellar Feedback With LIM} \label{sec:results:feedback}

In the left panel of Figure~\ref{fig:snapshots}, we show slices of $\delta T_b$ fluctuations in different feedback scenarios at various stages of reionization, when the average IGM neutral fraction is $\langle x_{\rm H \textsc{i}} \rangle = 0.2$, 0.5, and 0.8, respectively. It is obvious that at a given stage, the typical size of ionized regions is on average larger with stronger feedback. This is because under stronger feedback regulation, star formation tends to occur in more massive halos, which are more clustered and have a higher ionization rate to ionize a larger volume of gas thanks to their higher SFR. Note that although across each row the volume filling factor of fully ionized regions appears higher in the case of a stronger feedback, the volume-averaged neutral fraction $\langle x_{\rm H \textsc{i}} \rangle$ of individual simulated boxes are in fact comparable --- because the \textit{product} of ionization efficiency $\zeta$ and local collapse fraction $f_\mathrm{coll}(x, z, R)$ is more evenly distributed when feedback is less efficient, more partially ionized cells with ionized fraction equal to $\zeta^{-1} f_\mathrm{coll}(x, z, R_\mathrm{cell})$ are allowed to exist \citep{MFC_2011}, which compensate for the deficit in fully ionized regions. 

In Section~\ref{sec:results:skyavg}, we have demonstrated the impact of feedback on the history of cosmic dawn and reionization eras as revealed by the 21 cm global signal from the neutral IGM. Using LIMFAST, we supplement such a picture with the complementary LIM signals of UV/optical and far-infrared nebular emission lines tracing star-forming galaxies, which are considered to provide the majority of ionizing photons required to complete the reionization by $z\approx5.5$ \citep{Robertson_2015, Naidu_2020}. 

The right panel of Figure~\ref{fig:snapshots} shows slices through the boxes of H$\alpha$, Ly$\alpha$, and [\ion{C}{2}] intensity fluctuations simulated by LIMFAST when $\langle x_{\rm H \textsc{i}} \rangle \approx 0.5$ in different feedback scenarios, color-coded by the logarithmic line intensity in units of $\mathrm{Jy/sr}$. In contrast to maps of the 21 cm signal, these line intensity maps generally trace the ionizing sources residing in overdense regions, whose spatial anti-correlation with the 21 cm signal is clearly visible on scales larger than the typical size of ionized regions. On finer scales, information about sources of line emission and the luminosity distribution of the source population is encoded in detailed features of the intensity fluctuations. For Ly$\alpha$, a spatially-extended component is apparent, especially in the case of a strong feedback, which is sourced by recombinations in the diffuse ionized IGM surrounding ionizing sources. We note that, as discussed in Paper~I, we present intrinsic line intensity fields throughout, without including the attenuation effect due to dust grains in the ISM or patches of neutral hydrogen in the IGM. Such effects can be readily incorporated via post-processing the simulation boxes that LIMFAST outputs, as have been demonstrated in the literature \citep{Silva_2013, Heneka_2017}. 

\begin{figure*}[ht]
 \centering
 \includegraphics[width=0.98\textwidth]{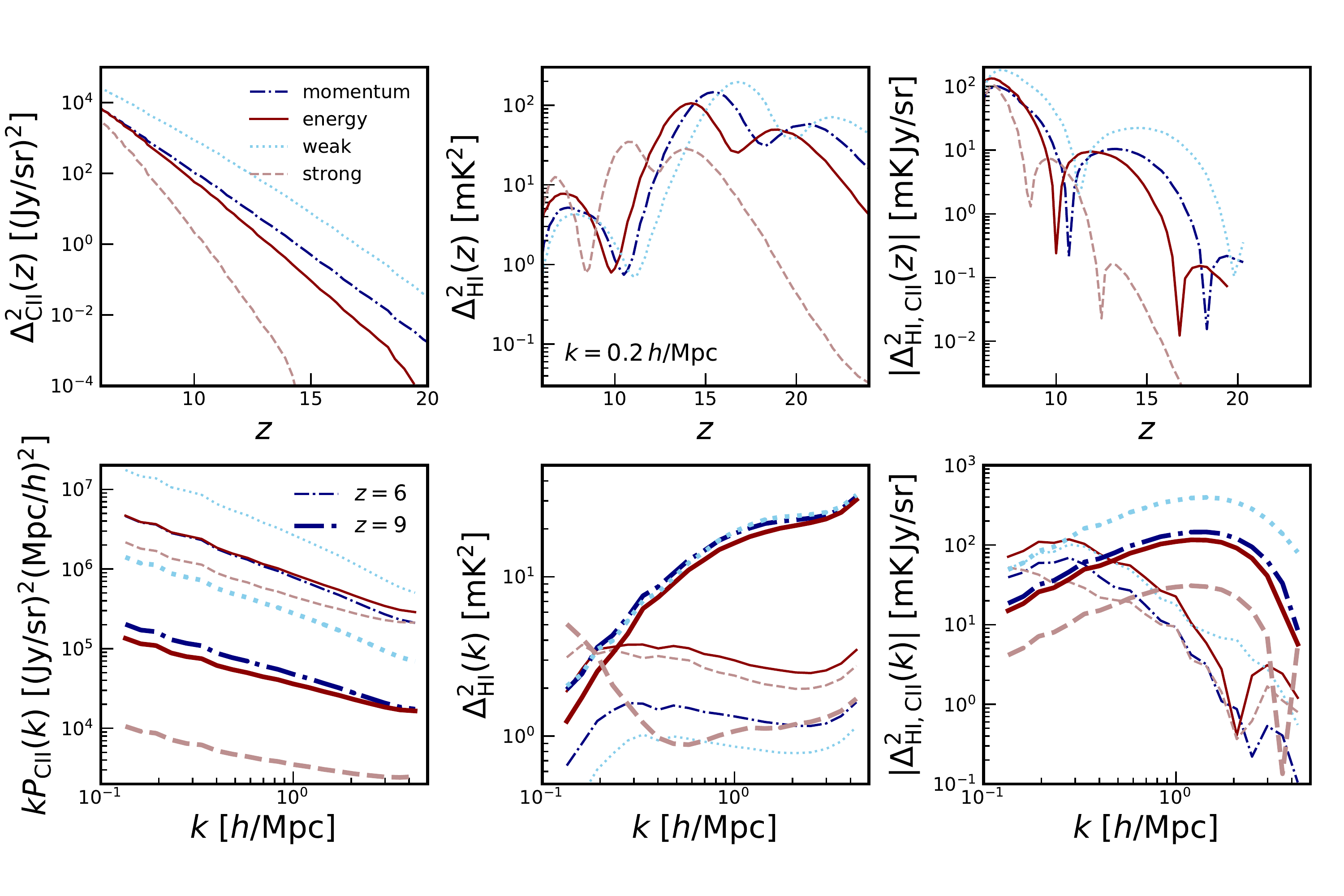}
 \caption{The redshift evolution at a fixed comoving scale of $k=0.2\,h/\mathrm{Mpc}$ (top) and scale dependence (bottom) of auto- and cross-power spectra (absolute value) between [\ion{C}{2}] emission from galaxies and the 21 cm line in different feedback models. The cross-power spectrum changes sign from negative on large scales to positive on small scales at $k_\mathrm{trans}\sim2\,h/\mathrm{Mpc}$. Note the different format in which the [\ion{C}{2}] auto-power spectra are plotted as a function of $k$ to facilitate visual comparison.}
 \label{fig:ps_feedback}
\end{figure*}

On the other hand, the fact that the [\ion{C}{2}] intensity field shows a larger spatial gradient compared with that of H$\alpha$ indicates that the former is preferentially sourced by more massive and therefore more biased sources. This results from the difference in the luminosity--halo mass ($L$--$M$) relation of the two lines. As demonstrated in Appendix~\ref{sec:lm_relation}, where we contrast the $L$--$M$ relation of several nebular lines under varying assumptions of the metallicity dependence, [\ion{C}{2}] luminosity is a steeper function of halo mass compared with lines like H$\alpha$ due to its much stronger metallicity dependence. The paucity of contribution from low-mass halos not only leads to a steep $\bar{I}(z)$ evolution shown in Figure~\ref{fig:skyavg}, but also implies that [\ion{C}{2}] intensity fluctuations will be more dominated by the Poisson noise from rare [\ion{C}{2}]-bright sources, as can be shown by the power spectrum. As we will see, a sensitive luminosity--halo mass relation makes statistical measurements of lines like [\ion{C}{2}] and CO promising ways of testing models of galaxy formation involving different feedback assumptions. 

\subsubsection{Information From Auto-Power Spectra} \label{sec:results:feedback:auto}

The statistical information about spatial fluctuations of a given LIM signal is directly available from its auto-power spectrum. As an example, we illustrate in Figure~\ref{fig:ps_feedback} the power spectra of 21 cm and [\ion{C}{2}] lines, whereas similar illustrations for the power spectra of other nebular lines considered (and their cross-correlations with the 21 cm line) are provided in Appendix~\ref{sec:more_lines}. The left two columns of Figure~\ref{fig:ps_feedback} show the shape and redshift evolution of [\ion{C}{2}] and \ion{H}{1} 21 cm power spectra calculated from simulations boxes in the four cases of feedback considered. Even though the power spectrum only partially describes these potentially highly non-gaussian fields, it is encouraging to see that useful information about the feedback mode in play can be probed by either the shape or amplitude evolution of the auto-power spectrum. 

From the redshift evolution of the power spectrum at $k=0.2\,h/\mathrm{Mpc}$ shown in the top row, it is evident that the amplitude of large-scale fluctuations encapsulates statistics of the key drivers for the sky-averaged signal evolution. In the case of 21 cm power spectrum amplitude, the three characteristic peaks (from high $z$ to low $z$) corresponds to the eras of Ly$\alpha$ coupling, X-ray heating, and reionization, when high-amplitude fluctuations in $\delta T_{\rm b}$ are concurrent with rapid changes (i.e., steep slopes) in the 21 cm global signal as shown in Figure~\ref{fig:skyavg}. Different feedback prescriptions modulate these peaks in significantly different ways, with strong feedback yielding peaks later and more squeezed in redshift and less contrasted in amplitude. 

The redshift evolution of [\ion{C}{2}] power spectrum amplitude, on the other hand, largely reflects the sky-averaged intensity $\bar{I}(z)$ evolution of the signal, which in turn traces the cosmic SFRD evolution as discussed in Section~\ref{sec:results:skyavg}. Thanks to the quadratic dependence $\Delta^2(k) \propto \bar{I}^2$, different feedback modes become more distinguishable, provided that the power spectrum amplitude can be monitored over a wide enough redshift range. 

From the shape of power spectra at $z=6$ and 9 shown in the bottom row, it is also straightforward to see the modulation effect by feedback. For 21 cm power spectrum, the impact of feedback on the ionized bubble size is manifested by the shift of the scale at which the power spectrum peaks, which is most discernible at $z=6$ ($k\sim0.1\,h/\mathrm{Mpc}$ for the ``strong'' model and $k\sim0.4\,h/\mathrm{Mpc}$ for the ``weak'' model) when different feedback models predict similar $\langle x_{\rm H \textsc{i}} \rangle$ but different BSDs. The difference appears to be a lot smaller at $z=9$ when the remains close to high, except for the case of very strong feedback which shows a qualitative difference from other cases. This is because the 21 cm spin temperature field is dominated by highly-biased, massive sources in the presence of strong feedback, thereby showing a distinctive large-scale power excess at $k\sim0.1\,h/\mathrm{Mpc}$ due to the source clustering. The generally much lower amplitude on smaller scales in case, compared with other feedback cases, is due to the delayed reionization by strongly-suppressed cosmic star formation. 

The shape evolution of [\ion{C}{2}] power spectrum is much more subtle in the plot, although the effect of feedback can still be inferred from the shape contrast between two redshifts. Overall, stronger feedback leads to both a steeper $\Delta^2_\mathrm{CII}(k)$ that is more dominated by the small-scale Poisson noise and a stronger shape evolution. Considering a metric of the power spectrum shape contrast, $\mathcal{X}(k_1, z_1, k_2, z_2)=\Delta^2(k_1, z_2)/\Delta^2(k_2, z_2) - \Delta^2(k_1, z_1)/\Delta^2(k_2, z_1)$, which characterizes the change in the dominance of small-scale Poisson noise in the power spectrum between two redshifts $z_1$ and $z_2$, we find $\mathcal{X}_\mathrm{CII}(3\,h/\mathrm{Mpc}, 6, 0.1\,h/\mathrm{Mpc}, 9)=20$, 40, 65, and 140 for the ``weak'', ``momentum'', ``energy'', and ``strong'' models, respectively. Here $k=3$ and $0.1\,h/\mathrm{Mpc}$ roughly correspond to the smallest and largest scales accessed by our simulation, and a larger, positive $\mathcal{X}$ indicates that from $z=6$ to 9 the ``strong'' model implies a larger increase in the dominance of the Poisson noise contribution. Such a correlation between $\mathcal{X}$ and the feedback strength is insensitive to factors that only affect the power spectrum normalization, and thus marks a potentially useful application of auto-correlation analysis to the understanding of galaxy formation physics. 

\subsubsection{Information From Cross-Power Spectra} \label{sec:results:feedback:cross}

In practice, measurements of the auto-power spectrum are often unfortunately complicated by a variety of astrophysical and instrumental effects. One of the main obstacles is foreground contamination, which can overwhelm the target LIM signal by several orders of magnitude. Even though a multitude of cleaning techniques have been devised to remove foreground contamination of various origins, cross-correlating signals with uncorrelated foregrounds still has its unique advantages. Therefore, it is interesting to understand how cross-correlations between different lines, especially the 21 cm line and nebular lines tracing ionizing sources, may be leveraged to characterize the effect of feedback in high-$z$ galaxy formation.

In the rightmost column of Figure~\ref{fig:ps_feedback}, we compare different feedback models by showing how their predicted 21 cm--[\ion{C}{2}] cross-power spectra evolve with redshift in their amplitude and shape. From the redshift evolution, the strength of feedback determines how rapidly the cross-power amplitude evolves. Moreover, thanks to the counteractive evolution of 21 cm and [\ion{C}{2}] amplitudes with redshift, the peaks intrinsic to the 21 cm contribution become broadened for $\Delta^2_{\rm H\,\textsc{i}, C\,\textsc{ii}}(z)$ compared with $\Delta^2_{\rm H\,\textsc{i}}(z)$ (especially for the peak due to X-ray heating), and the extent of the broadening depends on how long the counteractive effect persists. Different feedback modes are therefore easier to be distinguished by $\Delta^2_{\rm H\,\textsc{i}, C\,\textsc{ii}}(z)$, 
wherein less rapid evolution with broad, flattened, and later peaks in redshift corresponding to weaker feedback. From the shape of the cross-power spectrum, on the other hand, the most pronounced feature is the dependence of the scale at which the cross-power changes sign on the feedback mode in play. As will be shown next, both feedback and the physics of line emission affect the interpretation of such a characteristic scale, which has been perceived as an indicator of the typical size of ionized bubbles during the EoR. 

\begin{figure}[!ht]
 \centering
 \includegraphics[width=0.47\textwidth]{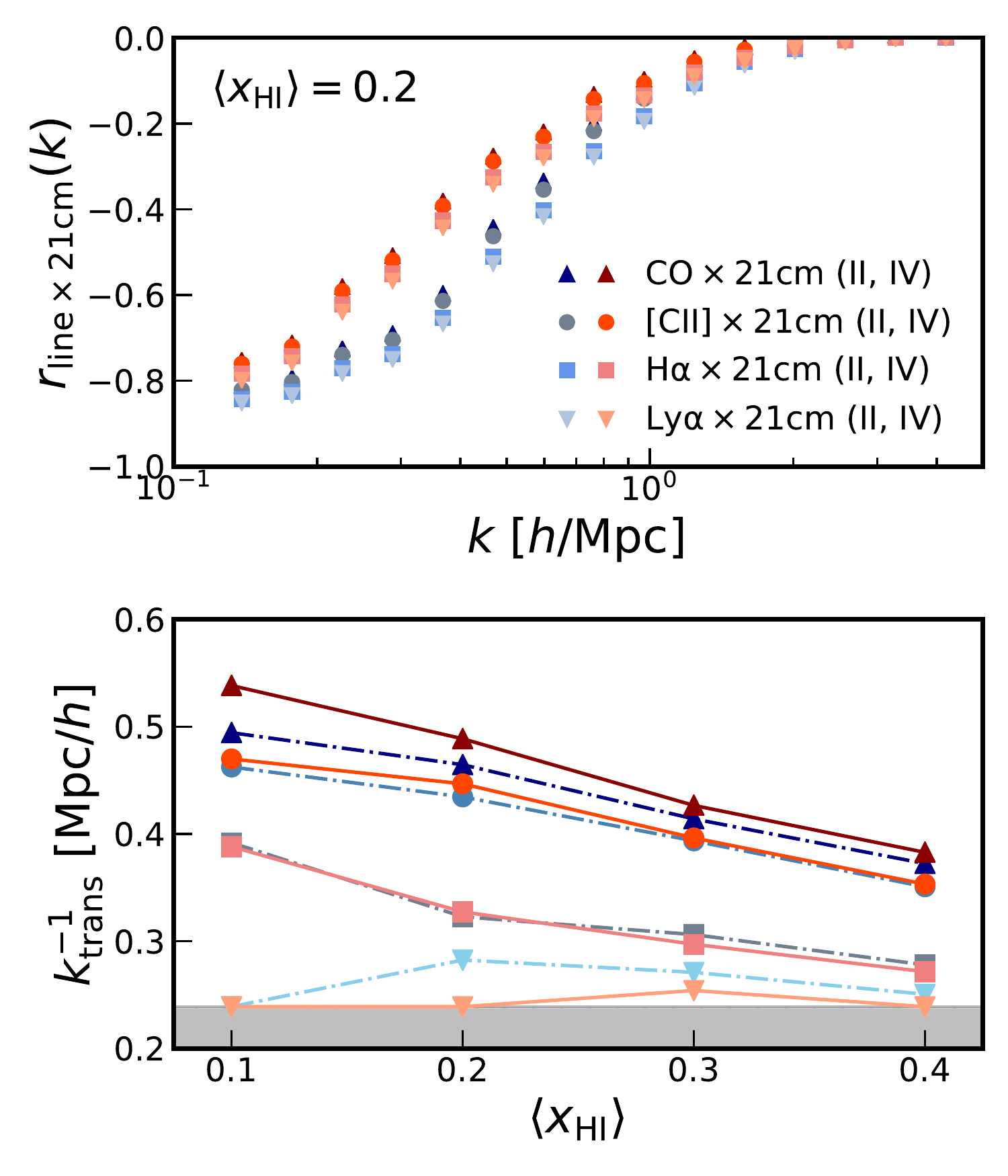}
 \caption{Top: cross-correlation coefficients between the 21 cm line and H$\alpha$, Ly$\alpha$, [\ion{C}{2}], and CO lines derived from the maps simulated by LIMFAST at $\langle x_{\rm H \textsc{i}} \rangle \approx 0.2$, assuming energy-driven (Model~II, blue set of points) and strong (Model~IV, red set of points) feedback. Bottom: the evolution of the comoving transition scale, at which $r_\mathrm{line \times 21cm}(k_\mathrm{trans})=0$, with the mean IGM neutral fraction. Scales inaccessible by our simulation outputs of limited resolution are greyed out, and $k^{-1}_\mathrm{trans}$ values below which are marked and interpreted as upper limits.}
 \label{fig:ccc_fb}
\end{figure}

We further inspect the effect of feedback on the cross-correlation signals in Figure~\ref{fig:ccc_fb} by comparing the cross-correlation coefficient $r_{1\times2}(k) = P_\mathrm{1\times2}(k)/\sqrt{P_1(k) P_2(k)}$ of the 21 cm signal with a variety of emission-line tracers of galaxies, including H$\alpha$, Ly$\alpha$, [\ion{C}{2}], and CO lines. In particular, we focus on how the scale dependence of $r(k)$ differs for different cross-correlations assuming different feedback assumptions, especially the transition scale $k_\mathrm{trans}$ where $r(k_\mathrm{trans})=0$. In the top panel of Figure~\ref{fig:ccc_fb}, $r(k)$ of different feedback models and line tracers when $\langle x_{\rm H \textsc{i}} \rangle \approx 0.2$ are shown in different hues and tints, respectively. The fact that, at a fixed stage (i.e., $\langle x_{\rm H \textsc{i}} \rangle$) of reionization, stronger feedback predicts faster de-correlation between 21 cm and nebular lines as $k$ increases (e.g., from $0.1\,h/\mathrm{Mpc}$ to $1\,h/\mathrm{Mpc}$) is consistent with the more ``top-heavy'' BSD skewed towards larger bubble sizes expected in this case. Nonetheless, in the bottom panel of Figure~\ref{fig:ccc_fb}, we show that $k_\mathrm{trans}$ is only modestly sensitive to feedback (and thus the BSD) for a given line tracer, although, as expected, it indeed traces the macroscopic progress of reionization described by $\langle x_{\rm H \textsc{i}} \rangle$. 

\begin{figure*}
 \centering
 \includegraphics[width=0.95\textwidth]{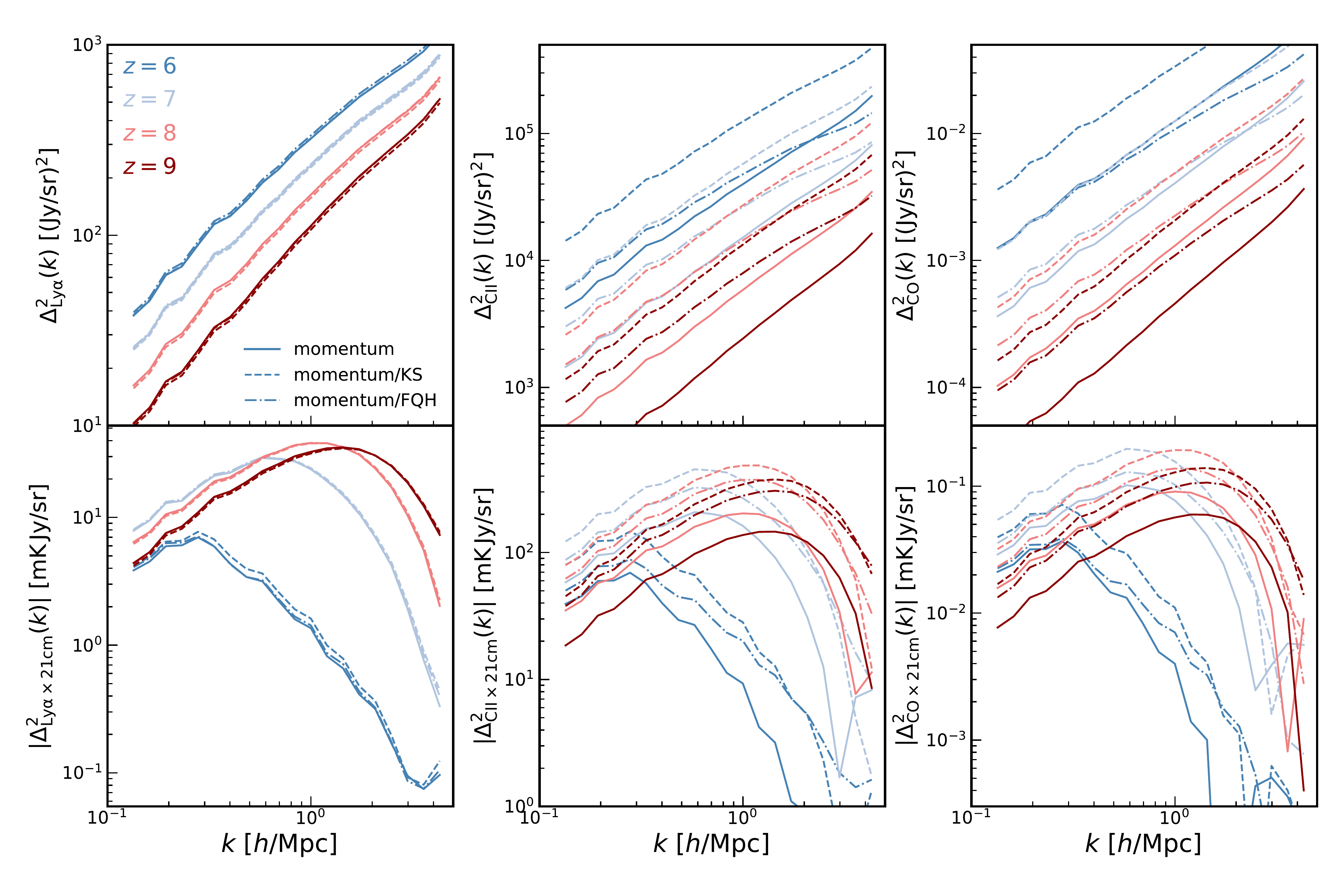}
 \caption{Power spectra of auto-correlations (top row) of Ly$\alpha$, [\ion{C}{2}], and CO lines, as well as their cross-correlations with the \ion{H}{1} 21 cm signal (bottom row) during the EoR predicted by the three LIMFAST models assuming the same momentum-driven feedback but different star formation laws.}
 \label{fig:ps_sflaw}
\end{figure*}

Another noteworthy feature in Figure~\ref{fig:ccc_fb} is the discrepancies in the individual cross-correlations for a given feedback mode. Unlike naively expected, there are non-trivial differences in both $r(k)$ and $k_\mathrm{trans}$ among different line tracers of galaxies. Lines like [\ion{C}{2}] and CO tracing the neutral ISM, whose luminosities evolve steeply with mass due to e.g., their strong dependence on the gas metallicity, exhibit a modestly lower level of (negative) correlation with the 21 cm line, when compared with tracers of the ionized ISM less sensitive to metallicity, such as H$\alpha$ and Ly$\alpha$. This, in turn, makes $k_\mathrm{trans}$ vary among different nebular lines with essentially different effective bias factors for a fixed feedback/reionization scenario. For example, $k_\mathrm{trans}^{-1}$, as a proxy for the bubble size, can differ by more than 50\% at $\langle x_{\rm H \textsc{i}} \rangle \sim 0.3$ depending on whether H$\alpha$ or CO is cross-correlated with the 21 cm line. Similar effects have been noted previously by several other authors \citep{Dumitru_2019, Kannan_2022_LIM, Cox_2022}, despite using less explicit formulations of the connection between nebular line emission and galaxy formation. Finally, for Ly$\alpha$, we note that a qualitatively different trend appears for $k_\mathrm{trans}^{-1}$ as a function of $\langle x_{\rm H \textsc{i}} \rangle$, which is caused by the additional diffuse component from recombinations in the diffuse ionized IGM (see Figure~\ref{fig:snapshots}) that can strongly modulate $k_\mathrm{trans}^{-1}$ especially towards the end of the EoR. 

\subsection{Characterizing the Star Formation Law With LIM} \label{sec:results:sflaw}

Besides stellar feedback, the other way that the astrophysics of galaxy formation can affect the luminosity--halo mass relation of nebular lines is through the star formation law. In particular, because the star formation law only alters the relative gas content of galaxies instead of the amount of star formation, as illustrated in Figure~\ref{fig:halo_growth}, lines originating from the neutral ISM are most sensitive to changes in the star formation law. We note, though, that measurements of large-scale structure using LIM signals of star-formation lines from \ion{H}{2} regions can still be useful probes of the star formation law across cosmic time \citep{Sun_2022}. 

Figure~\ref{fig:ps_sflaw} shows the auto-power spectra of Ly$\alpha$, [\ion{C}{2}], and CO lines and their cross-power spectra with the 21 cm signal during the EoR. Clearly, the statistics of Ly$\alpha$, whose luminosity simply scales with the star formation rate, are little affected by using different forms of the star formation law (Models~Ia, Ib, and Ic). Even though in principle there can be a small indirect effect through the different metallicities implied by different star formation laws, it is barely visible from the comparison of Ly$\alpha$ power spectra. On the contrary, given the substantial difference in the luminosity--halo mass (or SFR) relation caused by the gas mass dependence (see Figure~\ref{fig:lsfr}), [\ion{C}{2}] and CO lines have power spectra varying significantly with the assumed star formation law in both the shape and amplitude. Models implying more efficient star formation out of the gas reservoir and therefore a shallower luminosity--SFR relation, e.g., Model~Ic, tend to yield auto-power spectra less dominated by the Poisson noise and evolving less rapidly with redshift --- an unsurprising result given that large-scale fluctuations are mainly contributed by the more abundant fainter sources, whereas the small-scale Poisson fluctuations dominating are mainly contributed by the very rare and bright sources. For reference, we find $\mathcal{X}_\mathrm{CII}(3\,h/\mathrm{Mpc}, 6, 0.1\,h/\mathrm{Mpc}, 9)=41$, 25, 17, and $\mathcal{X}_\mathrm{CO}(3\,h/\mathrm{Mpc}, 6, 0.1\,h/\mathrm{Mpc}, 9)=55$, 36, 26 for Model~Ia, Ib, and Ic, respectively, suggesting that indeed Model~Ic assuming the FQH13 star formation law predicts power spectra the least Poisson noise-dominated. 

\begin{figure}[ht!]
 \centering
 \includegraphics[width=0.47\textwidth]{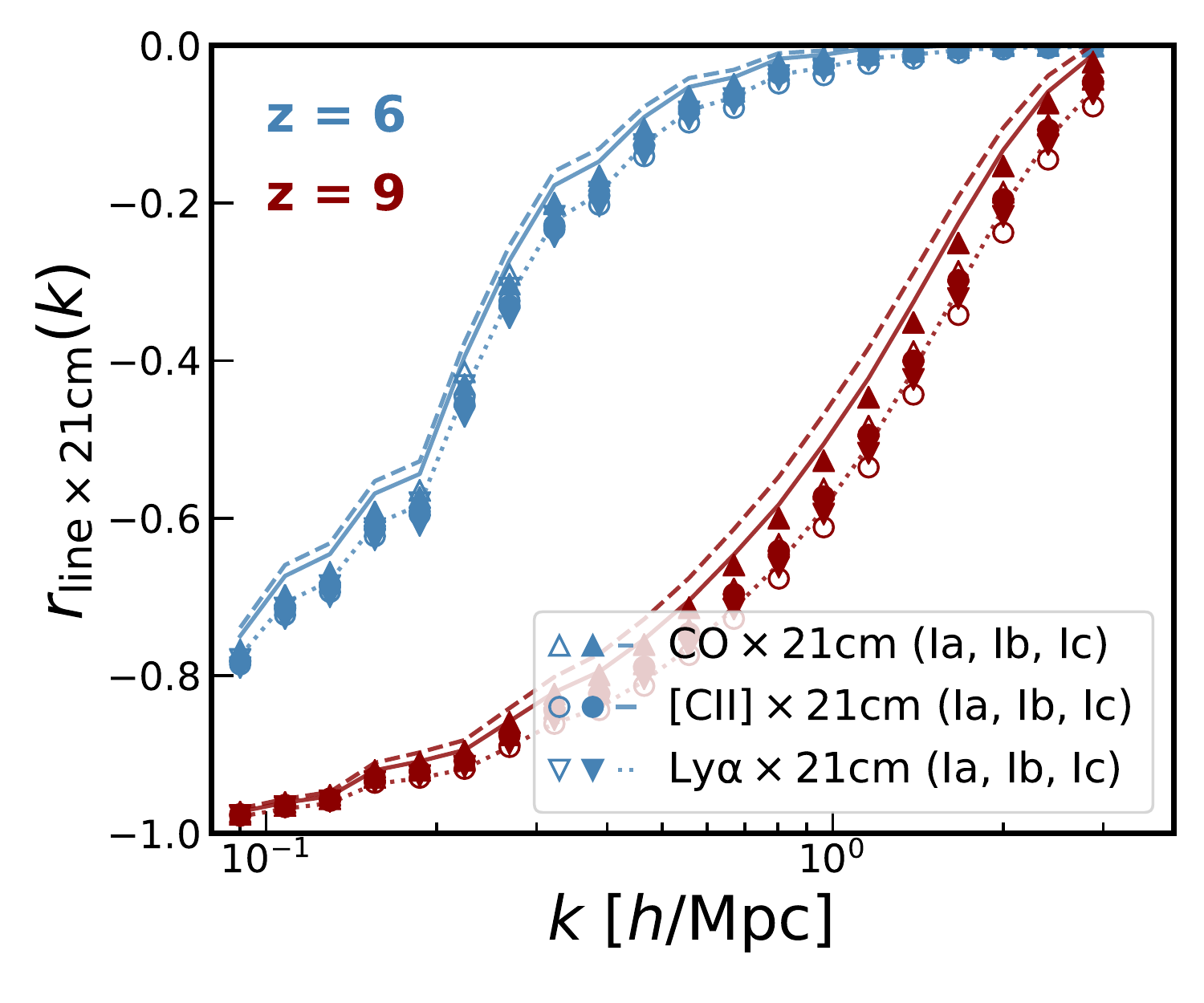}
 \caption{Cross-correlation coefficients between the 21 cm line and Ly$\alpha$, [\ion{C}{2}], and CO lines and derived from the maps simulated by LIMFAST assuming the default (lines, Model~Ia), KS (filled markers, Model~Ib), and FQH13 (empty markers, Model~Ic) star formation law.}
 \label{fig:ccc_sflaw}
\end{figure}

The cross-power spectrum between the 21 cm and [\ion{C}{2}] or CO lines also exhibits a clear dependence on the star formation law assumed, even though the reionization scenario is largely independent of it. With a steeper star formation law (i.e., more efficient gas to stellar mass conversion), the intensity field of [\ion{C}{2}] or CO becomes less dominated by bright sources and therefore de-correlates with the 21 cm field at smaller scales, causing a noticeable shape difference potentially useful for testing star formation law models. Moreover, the FQH13 model (Model~Ic) predicts the strongest redshift evolution of the cross-power amplitude over $6<z<9$, again due to the weaker counteractive evolution of 21 cm and [\ion{C}{2}] or CO lines in the case of a steeper star formation law.

In Figure~\ref{fig:ccc_sflaw}, we show the cross-correlation coefficients, $r(k)$, between Ly$\alpha$, [\ion{C}{2}], CO lines and the 21 cm line at $z=6$ and $z=9$ under various assumptions of the star formation law. Since the reionization scenario is nearly insensitive to changes in the star formation law, any difference in $r(k)$ shown in this figure is due to the nebular line intensity signal rather than the 21 cm signal. Several interesting features are noteworthy. First, as expected, $r_\mathrm{Ly\alpha \times 21cm}(k)$ remains almost unchanged in different star formation law models because Ly$\alpha$ only depends on the SFR. Second, similar to what is shown in Figure~\ref{fig:ccc_fb}, for either [\ion{C}{2}] or CO line the level of (negative) correlation at a given scale depends moderately on the star formation law assumed, with steeper star formation law yielding a less rapid de-correlation as $k$ increases. Lastly, a change in the relative order of $r(k)$ for Ly$\alpha$, [\ion{C}{2}], and CO lines is observed by contrasting Model~Ib with Model~Ic, which may be utilized for star formation law model selection. Similar to stellar feedback, the star formation law also serves as a source of complications in the interpretation of typical ionized bubble size from $r(k)$ or $k_\mathrm{trans}$. 

\section{Discussion} \label{sec:discuss}

In what follows, we compare the results and their implications from this work against some previous literature, and discuss potential caveats and limitations of our methods. In addition, we also outline several promising directions to extend the current framework of LIMFAST in the future. 

\subsection{Comparison to Previous Work}

A number of studies have previously studied and demonstrated the huge potential of LIM observations targeting at different tracers for understanding the cosmic dawn and reionization eras. By developing LIMFAST, we provide an efficient modeling framework to self-consistently simulate a large number of LIM signals during the EoR that have been investigated individually (or in small subsets) before by different authors, such as H$\alpha$ \cite[e.g.,][]{Heneka_2017, Silva_2018, Heneka_2021}, Ly$\alpha$ \cite[e.g.,][]{Silva_2013, Pullen_2014, Heneka_2017}, [\ion{C}{2}] \cite[e.g.,][]{Gong_2012, Chung_2020, Sun_2021a}, [\ion{O}{3}] \cite[e.g.,][]{Padmanabhan_2021}, and CO \cite[e.g.,][]{Lidz_2011, Mashian_2015, Breysse_2022}. As has been demonstrated in Paper~I, the heterogeneous assumptions made by different studies about physics of the ISM, star formation, feedback, and the metal and dust content often make direct comparisons between distinct line signals or distinct models of the same signal challenging and difficult to interpret. While qualitative comparisons may still reveal interesting astrophysical information, without quantitative assessments of the discrepancies observed it is unlikely to reliably test and compare different models against data. This urges the need to be able to describe and forecast various target LIM signals during the EoR --- usually differing in both the natal phase of gas and the connection to galaxy properties --- with a unified picture of high-$z$ galaxy formation. Nonetheless, as we show in Paper~I, non-trivial offsets exist between our results and other individual, line-specific models involving vastly varying assumptions of the galaxy population and spectral line production. Thus, coherently modeling the otherwise disconnected physical conditions of multiple emission lines and galaxy evolution, using tools like LIMFAST, is imperative to understand and exploit the multi-tracer LIM technique for studying the EoR. 

On the usage of the scale $k_\mathrm{trans}$ at which the cross-correlation coefficient between the 21 cm signal and a given galaxy tracer changes sign, our findings are qualitatively similar to previous analyses by \citet{Lidz_2011}, \citet{Dumitru_2019}, and most recently \citet{Kannan_2022_LIM}. Put briefly, the general redshift evolution of $k_\mathrm{trans}$ does reflect the overall progress of the reionization as measured by $\langle x_{\rm H \textsc{i}} \rangle$, but such evolution is complicated by uncertainties of the source population that affect signals of both tracers being cross-correlated. Specifically, at any given $\langle x_{\rm H \textsc{i}} \rangle$, variations of our galaxy model in either feedback or the star formation law can modulate $k_\mathrm{trans}$ through of the BSD and/or the effective bias of the galaxy tracer. While previous analyses often adopt a sharp dichotomy of halo emissivities in terms of ionizing photon production to distinguish between reionization scenarios dominated by faint vs. bright sources \cite[e.g.,][]{Dumitru_2019, Kannan_2022_LIM}, our model allows galaxies of different luminosities to more smoothly impact both the neutral gas and line intensity distributions in a consistent manner. Such smooth transitions in the contribution from different sources to signatures of reionization are not merely more realistic, but also essential for shedding light on how high-$z$ galaxies driving the reionization might be shaped by the balance between star formation and feedback. 

\subsection{Limitations of the Galaxy Formation Model} \label{sec:discuss:limitations}

In LIMFAST, we have implemented and leveraged the simple, quasi-equilibrium model of high-$z$ galaxy formation described in \citet{Furlanetto_2017} and \citet{Furlanetto_2021} to study the impact of the astrophysics of galaxies on various target LIM signals. Although it already represents an improvement over the source modeling in the latest release of 21cmFAST \cite[][]{Murray_2020} in aspects such as the physical connection between star formation and feedback regulation, some intrinsic limitations of the method need to be noted and are likely worthy of further exploration in future work. 

A key assumption made in our galaxy formation model is that in the high-$z$ universe a quasi-equilibrium state can already be established by proto-galaxies in the form of a settled disc where stars steadily form. Making this assumption provides a neat way to describe the formation of EoR galaxies by analogy to their low-$z$ counterparts, though one may question how valid such a scenario can be in the highly dynamic and uncertain stage of early galaxy formation. Recently studies, including a follow-up study to the \citet{Furlanetto_2021} model, have shown that star formation might be highly bursty during the early phase of galaxy formation, before some critical mass is reached and stars can steadily form. For example, \citet{FM_2022} generalize the quasi-equilibrium disc model by introducing a non-trivial perturbation arising from the time delay between star formation and stellar feedback at high redshifts. Numerical simulations also find strong evidence for strongly time-variable star formation in early, low-mass galaxies before a rotationally-supported ISM emerges from a rapid process of disc settling \citep{Gurvich_2022}, which turns out to be supported by Galactic archaeology of the in situ, metal-poor component of the Milky Way's stellar halo \citep{BK_2022}, indicating a potential requirement for full, non-equilibrium approaches. Given the intimate connection between star formation and spectral line emission in galaxies, as demonstrated in this work, it is crucial to quantify in future studies the effects of highly time-variable star formation on multi-tracer LIM observations of the EoR. 

Even if the quasi-equilibrium model indeed approximates the formation and evolution of high-$z$ star-forming galaxies well, it is admittedly simplistic in many ways, some of which are closely related to subgrid modeling that will be discussed in the next sub-section. One important simplification is associated with the diversity of galaxy formation histories. As demonstrated by \citet{Mirocha_2021}, simple subgrid, HAM-based models tend to produce biased signatures of the reionization process, when compared against fully numerical methods accounting for both halo mergers and the stochasticity of the halo mass accretion rate. A hybrid or numerically-calibrated approach will therefore be useful for further improvements in the model accuracy (see also Section~\ref{sec:discuss:extension}). Relatedly, we have also neglected the scatter in astrophysical parameters of our galaxy formation model, which can impact LIM signals of interest in a non-trivial way \citep{ShekharMurmu_2021, Reis_2022} and therefore should be taken into account in future development of LIMFAST by e.g., cell-level stochastic sampling of astrophysical parameters. Pop~III stars are another missing piece of the current model that can have non-trivial effects on the EoR, whose physical properties and formation histories may be studied either through their influence on the 21 cm signal \citep{Mirocha_2018, Mebane_2020, Qin_2021b, Munoz_2022} or by mapping the emission of nebular lines characteristic of Pop~III stars, such as the \ion{He}{2} 1640\,\AA\ line \citep{Visbal_2015, Parsons_2021}. While extensions of 21cmFAST-like, semi-numerical simulations have attempted to self-consistently model the formation of Pop~III and Pop~II stars altogether \cite[e.g.,][]{Tanaka_2021, Munoz_2022}, observational constraints, either direct or indirect, are pivotal to the down-selection of the poorly constrained model space \cite[see e.g.,][]{Mirocha_2018, Sun_2021m}. 

\subsection{Uncertainties With Subgrid Astrophysics} \label{sec:discuss:subgrid}

We note that a range of simplifications and model assumptions are made for the subgrid astrophysics of galaxy formation and evolution, which are essential for the application of LIMFAST to the EoR science, but in the meantime serve as important sources of uncertainty. For instance, the galaxy properties captured by our quasi-equilibrium model are highly simplified, which in turn limits how closely galaxy evolution and the production of the various kinds of spectral line emission can be modeled coherently. In particular, several physical conditions of the stellar population and the ISM must be specified manually, such as the star formation history of galaxies, the gas density, the interstellar radiation field strength, and the dust content and properties, all of which are likely influential for the modeling of both galaxy evolution and the LIM signals of interest \cite[e.g.,][]{Lagache2018, Mirocha_2020, Mirocha_2021, Yang_2021}. 

Given all the aforementioned uncertainties, as well as those mentioned in Section~\ref{sec:discuss:subgrid} about the galaxy formation model, accurately computing the line emission and eventually applying the model to reverse-engineer the properties of galaxies from upcoming cosmological surveys of the EoR will be a non-trivial task. Insights from observations and detailed numerical simulations of the mechanisms behind and the connections among different ingredients of subgrid astrophysics, such as the co-evolution of gas, metals, and dust across cosmic time \citep{Li_2019}, the connection between the ionization parameter and metallicity \citep{JY_2022}, and the effects of non-equilibrium photoionization and metal cooling \citep{Katz_2022}, will be extremely valuable for better understanding and further improvements of the source modeling in semi-numerical tools like LIMFAST, especially for the applications of interpreting future observations. 

\subsection{Extension of the Current Framework} \label{sec:discuss:extension}

In Paper~I and this work, we present the current structure and functionalities of LIMFAST focusing on its capability of forward modeling the multi-tracer LIM observations of the EoR. It is useful to note that the current framework may be readily extended in various promising ways and applied to a broader range of EoR studies, thanks to the modular nature of LIMFAST. 

First, additional probes of the EoR can be incorporated into the same modeling framework in a consistent manner similar to the existing ones. For instance, several authors have demonstrated that semi-numerical simulations are ideal tools for studying the kinetic Sunyaev-Zel'dovich (kSZ) effect from patchy reionization and its synergy with the 21 cm signal for constraining the reionization history \citep{Battaglia_2013, PLP_2020, Gorce_2022}. Cross-correlating the kSZ signal derived from the simulated ionization and velocity fields with line tracers of galaxies provides the redshift information missing in kSZ measurements. Similar ideas can be applied to other types broad-band, two-dimensional datasets such as the CMB lensing and the cosmic near-infrared background, through the large-scale fluctuations of which rich information about the population of ionizing sources may be extracted \citep{Helgason_2016, Maniyar_2022, Sun_2021m, Mirocha_2022}. Furthermore, as demonstrated already in the low-$z$ universe, three-dimensional Ly$\alpha$ forest tomography serves as a promising probe of the large-scale distribution of the neutral IGM \citep{Lee_2018, Newman_2020}, which can be ideally suited for studying the late stages of the reionization process by itself or in combination with LIM datasets \citep{Qin_2021a}. It is interesting to implement these additional observables into LIMFAST to quantitatively assess their potential for probing the EoR, especially when jointly analyzed with LIM observations, and examine methods required for overcoming observational challenges like foreground contamination \cite[e.g.,][]{Zhu_2018, Gagnon-Hartman_2021}. 

Besides taking into account extra probes of the EoR, it is also of interest to extend LIMFAST further into the post-reionization universe ($0<z<5$). Galaxies during this age of active assembly and evolution are not only interesting by themselves but also important witnesses of the impact of reionization on galaxy formation, which will be studied by a number of forthcoming LIM surveys of galaxies at low-to-intermediate redshift, such as COMAP \citep{Cleary_2021}, EXCLAIM \citep{Cataldo_2020SPIE}, SPHEREx \citep{Dore_2018}, and TIM \citep{Vieira_2020}. That said, despite showing great promise, the low-$z$ extension of LIMFAST faces two main challenges. First, at lower redshift, the halo occupation distribution (HOD) becomes more sophisticated due to the increased population of satellite galaxies \citep{Kravtsov_2004, Bhowmick_2018, Behroozi_2019}, and quenching becomes a more and more important process in galaxy formation and evolution \citep{Tal_2014, Brennan_2015, Donnari_2021}. Both factors call for more detailed subgrid models for the luminosity--halo mass relation. Meanwhile, accurately modeling the partitioning of mass into halos becomes more challenging at lower redshift due to the increased importance of halo mergers. LIMFAST inherits the formulation of large-scale structure and radiation field approximation from 21cmFAST, where the generation of halo source fields by a halo finding algorithm is bypassed. To properly account for halo merger histories in the low-$z$ extension, an explicit halo finding algorithm, with either an extended dynamic range to resolve small halos at the cooling threshold, or an enhanced subgrid modeling of halo source fields involving merger trees and a stochastic population of simulation cells with unresolved halos, will be required at the cost of extra RAM capacity and a slower speed \cite[see discussions in e.g.,][and references therein]{MF_2007, MFC_2011}.

\section{Conclusions} \label{sec:conclusions}

Using simulations generated by the LIMFAST code introduced in Paper~I, we have presented in this paper a unified picture of how the astrophysics of high-$z$ galaxy formation affect and therefore can be reveal by multi-tracer LIM observations of the EoR. We investigate the impact of different stellar feedback and star formation law prescriptions on a variety of signatures of reionization, including the 21 cm signal and LIM signals of nebular emission lines from the multi-phase ISM, such as H$\alpha$, Ly$\alpha$, [\ion{O}{3}], [\ion{C}{2}], and CO. Our main findings can be summarized as follows: 
\begin{enumerate}
    \item Because the cosmic star formation history is sensitive to feedback-regulated star formation in individual galaxies, the efficiency of stellar feedback directly impacts the history, geometry, and thereby the variety of observational signatures of the reionization process. On the other hand, besides a small indirect effect through the metallicity, the star formation law only affects tracers of the neutral ISM of galaxies as indirect probes of the reionization. 
    
    \item The redshift evolution of multiple sky-averaged line signals already serves as a useful probe of the astrophysics of high-$z$ galaxy formation. Timings of the extrema in the 21 cm global signal are tightly connected to the feedback efficiency through radiation fields scaling with the cosmic SFRD. Due to the strong metallicity dependence of metal cooling lines like [\ion{C}{2}], a comparison between their sky-averaged signal evolution and that of hydrogen lines like H$\alpha$ can inform the (cosmic mean) stellar feedback strength.
    
    \item Rich astrophysical information about the reionization and its driving sources can be extracted from the auto- and cross-power spectra of multi-tracer LIM data. Both feedback and the star formation law can modulate the shape and amplitude of power spectra and their evolution across cosmic time. Power spectral analyses combining multiple, complementary tracers therefore allows cross-checks and the separation of effects due to the reionization itself and those associated with galaxy formation and evolution. 
    
    \item The cross-correlation between the 21 cm line and a spectral line tracer of galaxies is particularly useful for tracing the overall progress of the EoR. However, even though the transition scale $k_\mathrm{trans}$ roughly probes the neutral fraction evolution, the exact interpretation and implications of the cross-correlation are subject to complications due to astrophysics of galaxy formation and the resulting properties of galaxies, and thus dependent on the specific tracer considered. Multi-tracer LIM makes it possible to better understand how LIM signals are influenced by astrophysical processes such as feedback and the star formation law, on which the usage of $k_\mathrm{trans}$ or the cross-correlation analysis in general is premised. 
 
 \item By accessing a larger fraction of the faint galaxy population than individual source detection, LIM surveys can use the inferred SFRD to offer more sensitive tests for processes central to galaxy formation like the stellar feedback. This makes LIM a highly complementary method for studying high-$z$ galaxy formation even in the era of new-generation telescopes.
\end{enumerate}

In summary, there is great potential for multi-tracer LIM to transform our understanding of cosmic reionization and the formation and evolution of high-$z$ galaxies that drive the reionization process. In spite of the various challenges that commonly exist in practice for different tracers, such as the mismatch of scales and issues of foreground contamination \cite[see the review e.g.,][]{LS_2020}, careful coordination and optimization for future multi-tracer synergies will eventually allow the invaluable astrophysical information to be extracted and applied to tests of the galaxy formation theory at high redshift. Reliable semi-numerical simulations like LIMFAST, in its current and future forms, are essential tools for accurately modeling and analyzing the vast amount of observational data to come. 

\begin{figure}
 \centering
 \includegraphics[width=0.47\textwidth]{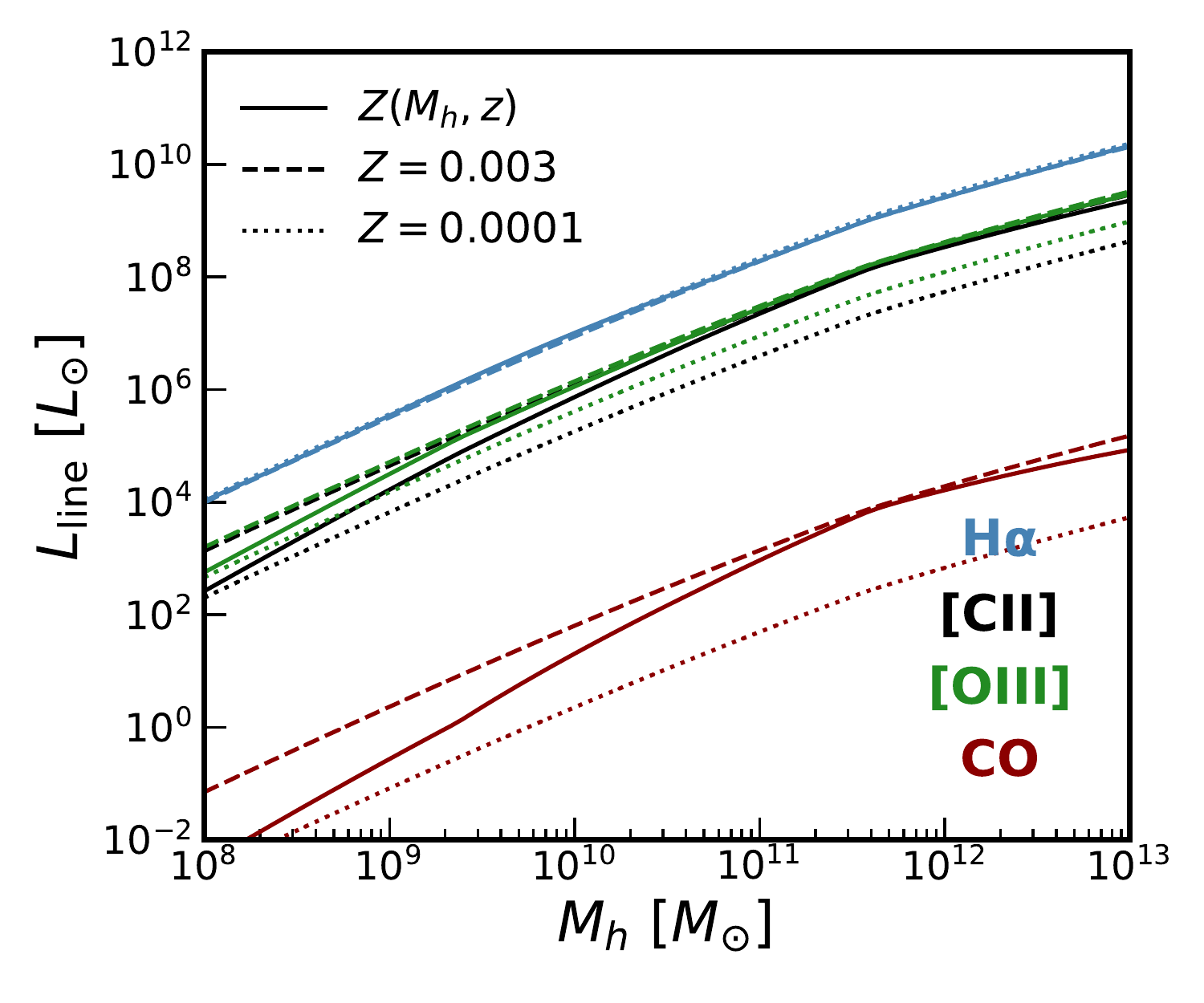}
 \caption{The $L$--$M$ relations of H$\alpha$, [\ion{C}{2}], [\ion{O}{3}], and CO(1--0) lines assuming different metallicity values. The solid curve assumes a varying metallicity predicted by our galaxy model (Model~Ia), whereas the dashed and dotted curves assume fixed metallicity values of $Z=0.003$ and $Z=0.0001$, respectively.}
 \label{fig:l_vs_m}
\end{figure}

\begin{figure*}
 \centering
 \includegraphics[width=0.49\textwidth]{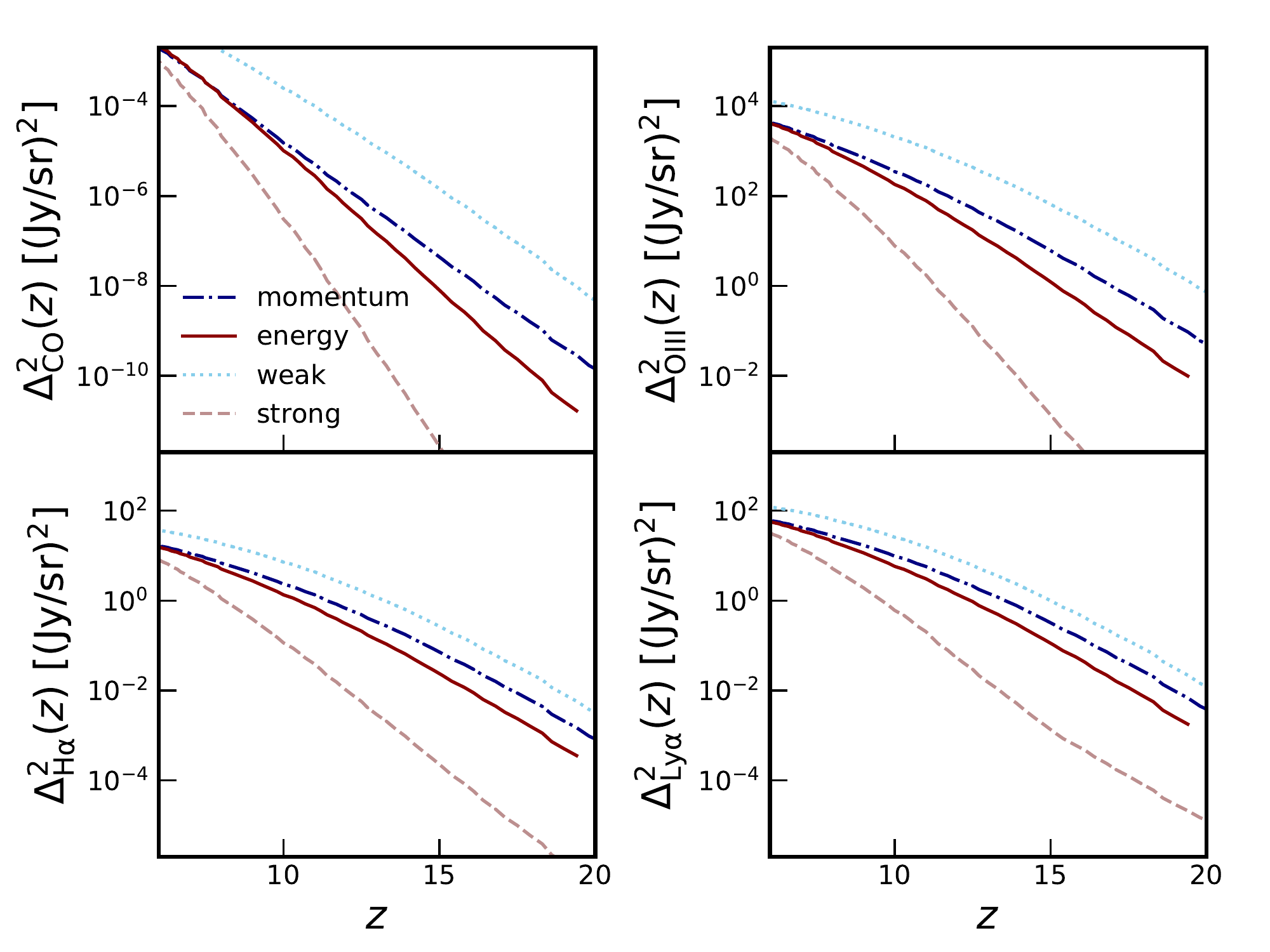}
 \includegraphics[width=0.49\textwidth]{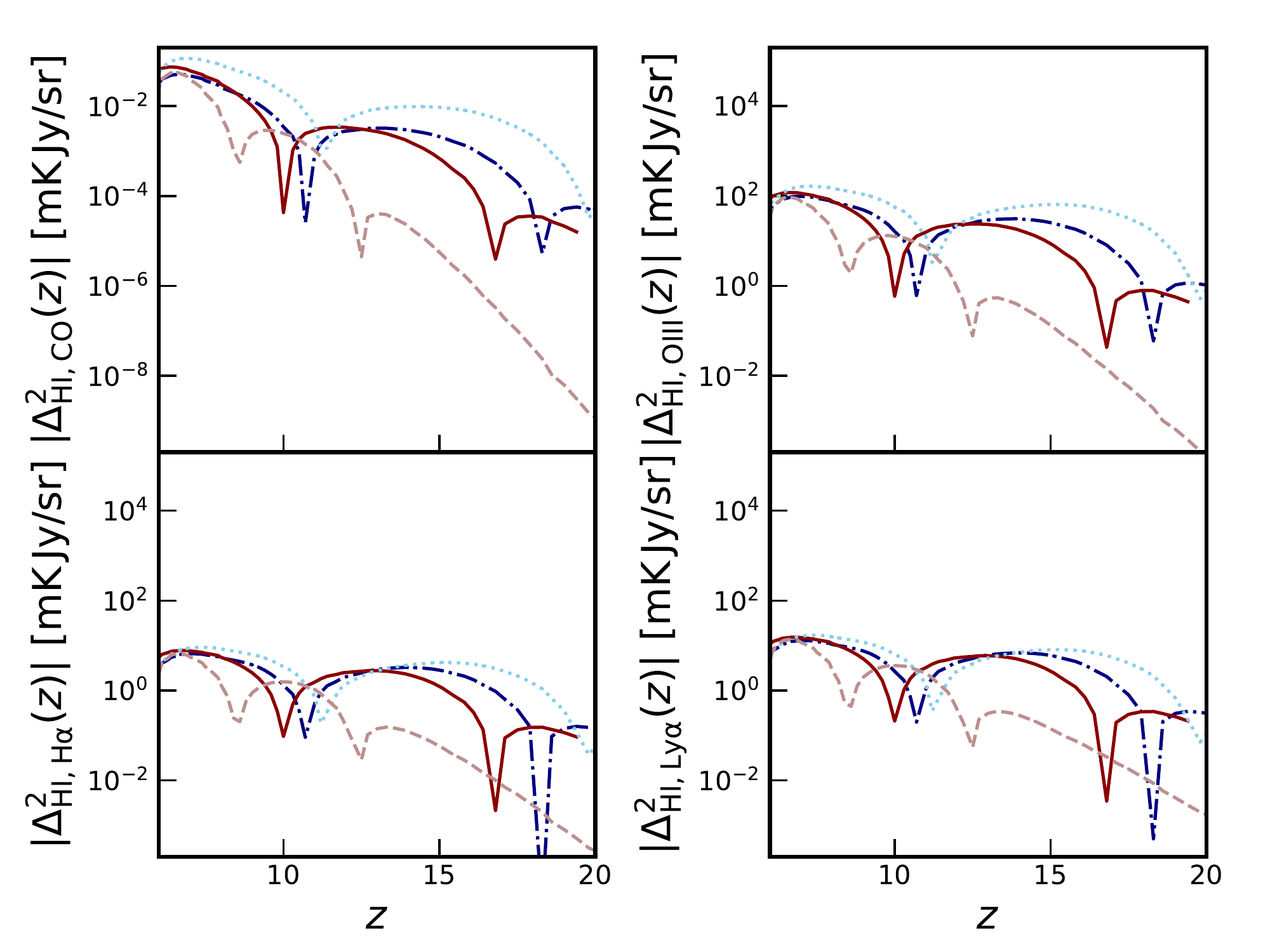}
 \includegraphics[width=0.49\textwidth]{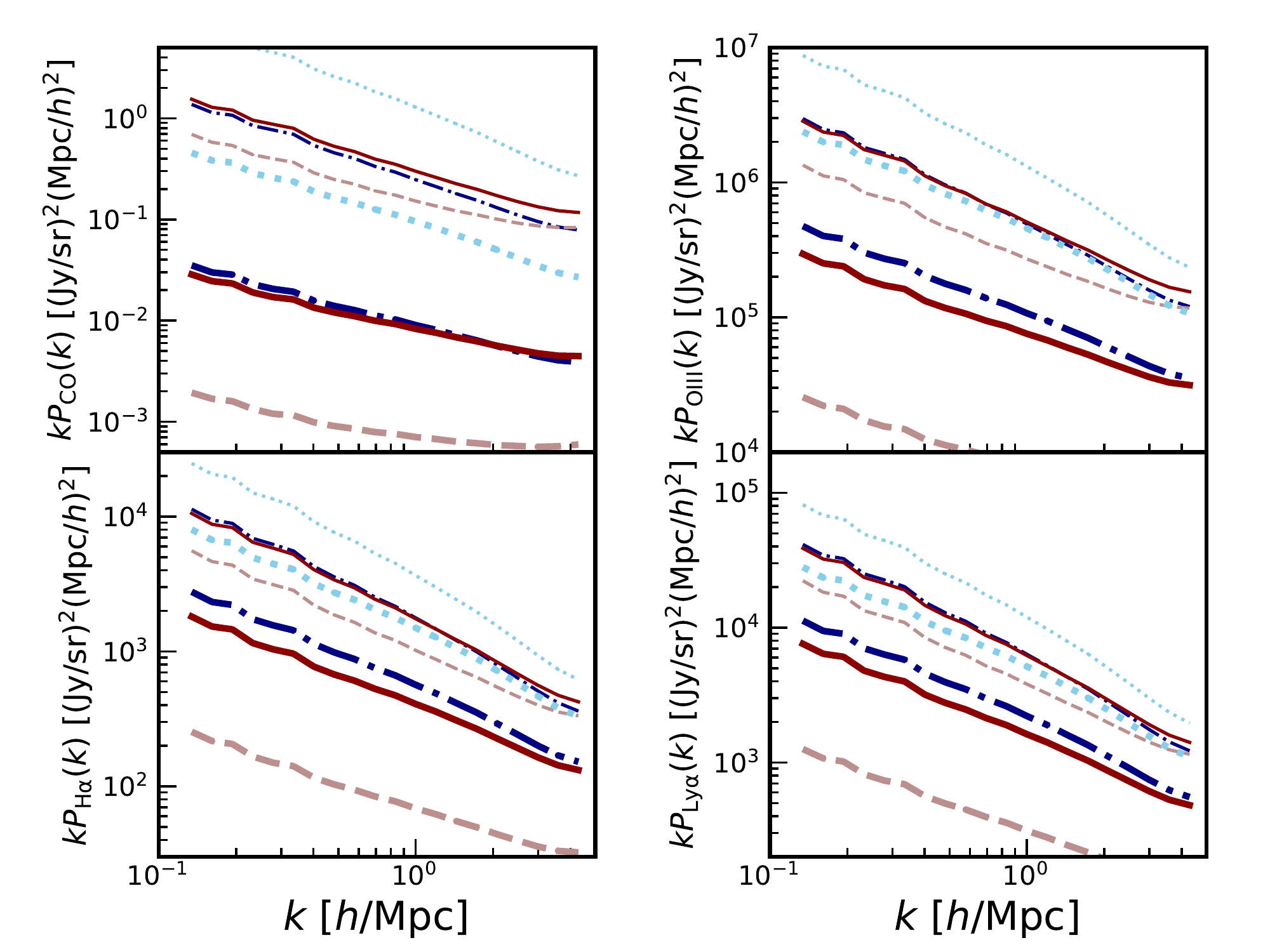}
 \includegraphics[width=0.49\textwidth]{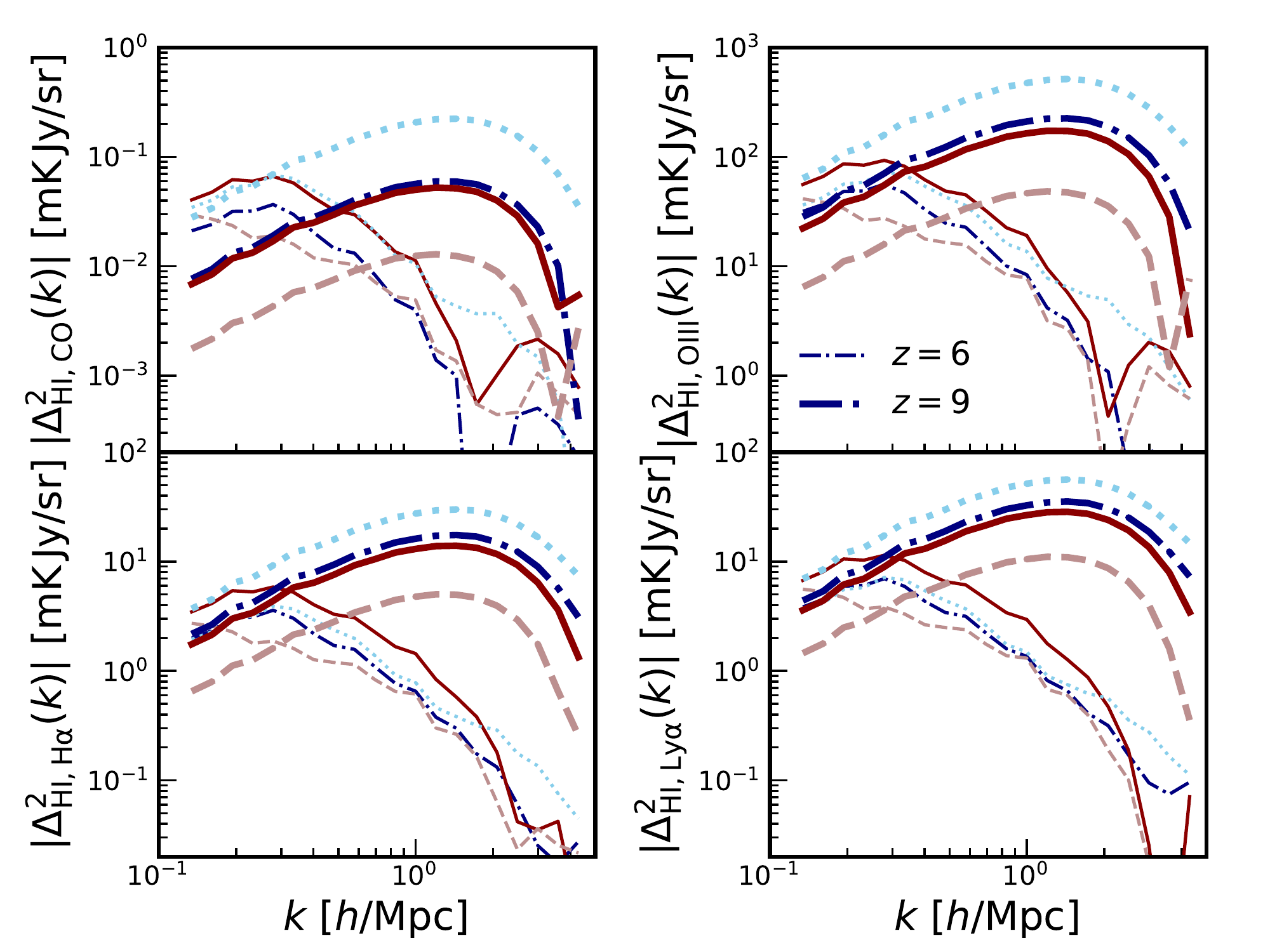}
 \caption{Same as Figure~\ref{fig:ps_feedback}, but for CO(1--0), [\ion{O}{3}], H$\alpha$, and Ly$\alpha$ lines. Note the different format in which the auto-power spectra are plotted as a function of $k$ to facilitate visual comparison.}
 \label{fig:ps_feedback_other_lines}
\end{figure*}

We thank Adam Lidz, Fred Davies, Jordan Mirocha, Rahul Kannan, and Yuxiang Qin for helpful discussions and comments that helped improve this paper, as well as Mauro Stefanon for sharing the data of the stellar-to-halo mass ratios. We acknowledge support from the JPL R\&TD strategic initiative grant on line intensity mapping. Part of this work was done at Jet Propulsion Laboratory, California Institute of Technology, under a contract with the National Aeronautics and Space Administration. 

\software{ARES \citep{Mirocha_2017}, BPASS \citep{Eldridge_2017}, \textsc{cloudy} \citep{Ferland_2017}, 21cmFAST \citep{MFC_2011}}

\appendix

\twocolumngrid

\section{Luminosity--Halo Mass Relation} \label{sec:lm_relation}

The $L$--$M$ relation directly dictates the way the underlying matter density field is traced by the line intensity map observed, as has been discussed in previous studies \cite[e.g.,][]{Kannan_2022_LIM}. Varying behaviors of the $L$--$M$ relation are therefore essential for understanding the properties and statistics of the spectral line tracer in different variations of the galaxy model. 

Figure~\ref{fig:l_vs_m} shows a comparison of the $L$--$M$ relation of different lines, assuming either a fixed metallicity or a varying metallicity as predicted by our galaxy model. The similar shapes of curves with a fixed metallicity suggests that the scaling with the SFR (e.g., H$\alpha$ and [\ion{O}{3}]) or the gas mass (e.g., [\ion{C}{2}] and CO) barely affects the $L$--$M$ relation in the case of the simple star formation law assumed for Model~Ia (and Models~II--IV). It is really the metallicity dependence and evolution that result in different $L$--$M$ relations of hydrogen and metal lines, which in turn lead to the different effective bias factors of these spectral line tracers. It is also interesting to note that the $L$--$M$ relation is coupled to the metallicity not only through the metal content of the ISM, but also through the metallicity dependence of the ISRF, whose synthetic spectrum from BPASS v2.1 \citep{Eldridge_2017} is supplied to the \textsc{\textsc{cloudy}} simulations (see Paper~I). For instance, a more ionizing ISRF for a lower metallicity produces modestly brighter H$\alpha$ emission, and partially counteracts the effect of a more metal-poor ISM towards lower halo masses for metal lines highly sensitive to the ionizing radiation like [\ion{O}{3}].

\section{Power Spectra of Nebular and 21 cm Lines With Varying Feedback} \label{sec:more_lines}

Supplementing Figure~\ref{fig:ps_feedback} which uses [\ion{C}{2}] and 21 cm lines to exemplify the ways feedback affects the redshift evolution and scale dependence of LIM power spectra, we further in Figure~\ref{fig:ps_feedback_other_lines} similar results for other nebular lines considered in this work, including [\ion{O}{3}], CO(1--0), H$\alpha$, and Ly$\alpha$ (sum of star formation and IGM contributions). Comparing the evolution with redshift and scale of different tracers shows interesting (though in some cases subtle) trends that inform about how these lines are sensitive to different aspects of the galaxy evolution such as the gas content and metallicity, which may be systematically probed by combining LIM surveys of all these different line tracers. 

\newpage

\bibliography{limfast_paper2}{}
\bibliographystyle{aasjournal}



\end{document}